\documentclass[a4,usenatbib]{mnras}
\usepackage{amsmath}
\usepackage{amssymb} 
\usepackage{multirow}
\usepackage{textcomp}
\usepackage{microtype}
\usepackage{subcaption}
\usepackage{xcolor,ulem}
\usepackage{graphicx}
\usepackage{rotating}
\usepackage{float}
\usepackage{mathptmx}
\usepackage{comment}
\usepackage{physics}

\newcommand{\email}[1]{\mbox{\href{mailto:#1}{#1}}}

\def\apj{Astrophysical Journal}                
\def\apjl{Astrophysical Journal}             

\def\apjs{ApJS}              
\def\mnras{MNRAS}            
\def\prd{Phys.~Rev.~D}       
\def\araa{ARA\&A}             
\def\aap{A\&A}                
\def\sovast{Soviet Ast.}

\title[Mass-redshift dependency of SMBHBs for the GWB]{Mass-redshift dependency of Supermassive Black Hole Binaries for the Gravitational Wave Background}

\author[Kozhikkal et al.]{\parbox{\textwidth}{
Musfar Muhamed Kozhikkal,$^1$\thanks{\email{musfarmuhamed@gmail.com}}
Siyuan Chen,$^{2}$\thanks{\email{sychen@pku.edu.cn}}
Gilles Theureau,$^{1,3,4}$
Mélanie Habouzit,$^{5,6}$
and Alberto Sesana$^{7,8,9}$}
\vspace{0.4cm}
\\
  $^1$Laboratoire de Physique et Chimie de l'Environnement et de l'Espace, Universit\'e d'Orl\'eans / CNRS, 45071 Orl\'eans Cedex 02, France\\
  $^2$Kavli Institute for Astronomy and Astrophysics, Peking University, Beijing 100871, P. R. China \\
  $^3$Observatoire Radioastronomique de Nan\c{c}ay, Observatoire de Paris, Universit\'e PSL, Université d'Orl\'eans, CNRS, 18330 Nan\c{c}ay, France\\
  $^4$Laboratoire Univers et Th{\'e}ories LUTh, Observatoire de Paris, Universit{\'e} PSL, CNRS, Universit{\'e} de Paris, 92190 Meudon, France\\
  $^5$Zentrum f\"{u}r Astronomie der Universit\"{a}t Heidelberg, ITA, Albert-Ueberle-Str. 2, D-69120 Heidelberg, Germany\\
  $^6$Max-Planck-Institut f\"{u}r Astronomie, K\"{o}nigstuhl 17, D-69117 Heidelberg, Germany\\ 
  $^7$Dipartimento di Fisica ``G. Occhialini", Universit{\'a} degli Studi di Milano-Bicocca, Piazza della Scienza 3, I-20126 Milano, Italy\\
  $^8$INFN, Sezione di Milano-Bicocca, Piazza della Scienza 3, I-20126 Milano, Italy\\
  $^9$INAF - Osservatorio Astronomico di Brera, via Brera 20, I-20121 Milano, Italy
  }

\date{Accepted XXX. Received YYY; in original form ZZZ}

\pubyear{2023}

\begin{document}
\label{firstpage}
\pagerange{\pageref{firstpage}--\pageref{lastpage}}
\maketitle

\begin{abstract}
  Studying how the black hole (BH) - (galaxy) bulge mass relation evolves with redshift provides valuable insights into the co-evolution of supermassive black holes and their host galaxies.
  However, obtaining accurate measurement of BH masses is challenging due to the bias towards the most massive and luminous galaxies.
  Instead we focus on the BH and bulge masses as they vary with redshift using the EAGLE, Illustris, TNG100, TNG300, Horizon-AGN and SIMBA large-scale cosmological simulations.
  We use an analytical astrophysical model with galaxy stellar mass function, pair fraction, merger timescale and BH-bulge mass relation extended to include redshift evolution.
  The model can predict the intensity of the gravitational wave background (GWB) produced by a population of supermassive black hole binary (SMBHB) as a function of the frequency.
  This allows us to compare the predictions of this model with the constraints of Pulsar Timing Array observations. Here, we employ Bayesian analysis for the parameter inference.
  We find that all six simulations are consistent $\leq 3.5\sigma$ with a range of simulated GWB spectra.
  By fixing the BH-bulge mass parameters to the simulations we analyze the changes in the constraints on the other astrophysical parameters. Furthermore, we also examine the variation in SMBHB merger rate with mass and redshift between these large-scale simulations.
\end{abstract}


\section{Introduction}
\label{sec:Introduction}

The co-evolution of galaxies and their supermassive black holes (SMBHs), i.e. the relationship between SMBHs and the dark matter halo potential, their role in the stellar formation activity, their local interactions with the stars and gas, and their fate during the history of galaxy mergers, are key ingredients of recent large cosmological simulations and of our understanding of large-scale structure formation and evolution \citep[see e.g.,][and references therein]{2021MNRAS.503.1940H,2022MNRAS.509.3015H}.

Moreover, the SMBH pair formation process in the post-merger galaxy potential and their inspiral to coalescence, produces gravitational waves (GWs) in the low frequency domain, observable either as a stochastic gravitational wave background (GWB) or as individual continuous gravitational wave sources with Pulsar Timing Array (PTA) experiments (nHz$-\mu$Hz) \citep{1990ApJ...361..300F,1995ApJ...446..543R,2003ApJ...583..616J,SesanaVecchioColacino:2008}, or with the future spatial laser interferometers like LISA ($10^{-4}$Hz$-10^{-2}$Hz) \citep{2017arXiv170200786A,2023LRR....26....2A}.

A PTA uses radio telescopes to time a network of millisecond pulsars \citep{Sazhin:1978,Detweiler:1979}. In principle, once the pulsar rotation irregularities, its possible orbital motion, the dispersion and scattering of its radio signal through the interstellar and heliospheric plasma and the systematics due to the Earth's motion in the Solar System are properly modelled and subtracted from the time series of measured pulsations, one expects to be able to extract the GW imprint from the resulting timing residuals. 
The analysis requires observations of multiple millisecond pulsars at sub $\mu$s precision for several decades (up to about 25 years for ongoing programs) in order to extract a GWB from unmodelled noise.
There are several PTA consortia, structured at continental levels and collaborating globally: European PTA (EPTA)\citep{2013CQGra..30v4009K,2016MNRAS.458.3341D,2021MNRAS.508.4970C}, Parkes PTA (PPTA) in Australia \citep{2013PASA...30...17M,2013CQGra..30v4007H,2020PASA...37...20K}, North American Nanohertz Observatory for Gravitational Waves (NANOGrav) \citep{2016ApJ...821...13A,2018ApJ...859...47A,2020ApJ...905L..34A},
Indian PTA (InPTA) \citep{2018JApA...39...51J,2022PASA...39...53T}, Chinese PTA (CPTA) \citep{CPTA,jyg+19} 
and MeerTime PTA (MPTA) in South Africa \citep{2020PASA...37...28B,2022PASA...39...27S}. These PTAs form a world wide organisation, the International PTA (IPTA), where they share their data and coordinate their analysis to eventually detect and hopefully characterise the GW signal \citep{2010CQGra..27h4013H,2016IPTA,2022MNRAS.510.4873A}.

NANOGrav, PPTA, EPTA and IPTA have reported the detection of a low-frequency common signal in their pulsar datasets~\citep{2020ApJ...905L..34A,2021ApJ...917L..19G,2021MNRAS.508.4970C,2022MNRAS.510.4873A}.
This marks the first step towards the detection of a GWB. If the common signal is of gravitational wave origin it should also show a characteristic spatial correlation between the pulsars, called the Hellings-Downs correlation \citep{HellingsDowns:1983}, 
which the above mentioned collaborations and the Indian Pulsar Timing Array have found evidence for \citep{eptadr2_gwb,ng15_gwb,pptadr3_gwb,ipta_comp}. In addition, the Chinese Pulsar Timing Array concurrently also found significant evidence for a Hellings-Downs correlated signal in their dataset \citep{cpta_dr1}.

If these recently observed spectral signatures are from a population of supermassive black hole binaries (SMBHBs), they favour heavy black hole masses and short merger timescales. Future detections will improve on these constraints and should allow some relations to be ruled out, in particular those with the lowest GWB. This would open new multi-messenger probes to study SMBHs and their host galaxies \citep[e.g.,][]{2021ApJ...911L..34P}.

By formulating the relative strength of the GWB as a function of SMBHB merger rate and gravitational wave energy spectrum, we can connect them to astrophysical parameters. The SMBHB merger rate is linked to the galaxy merger rate via a mass relation between the SMBH and galaxy bulge. Using the Galaxy Stellar Mass Function (GSMF), a differential pair fraction of galaxy in binaries and a merger timescale one can compute the galaxy merger rate. The gravitational wave energy spectrum depends on the binary orbital eccentricity and the nature of the environment driving their evolution \citep{2017MNRAS.470.1738C,2019MNRAS.488..401C}.  

The mass relation between the SMBH and galaxy bulge, called the BH-bulge mass relation, is widely studied using both observational data and large-scale cosmological simulations. The different values of the BH-bulge mass parameters for our Universe are constrained using observational data. Although there is currently no consensus, several observational samples suggest that the BH-bulge mass relation could evolve with redshift \citep{2010ApJ...708..137M,2013ARAA..51..511K}. In these papers, for a fixed galaxy mass, BHs are on average more massive at high redshift compared to those in similar host galaxies at low redshift.

Studying the evolution of the Universe through observations is a challenging task due to a number of technical limitations. 
The expansion of the Universe causes the light from the galaxies and SMBHs to shift towards longer wavelengths, making it difficult to detect their emission and accurately measure their properties, such as their mass and accretion rate. For example, it can be difficult to study scaling relations at high redshifts beyond $z\sim 2$ due to the challenges of disentangling the light from an active galactic nucleus (AGN) and the light from the host galaxy \citep{2020ApJ...888...37D}. The high redshift galaxies are fainter and smaller than nearby galaxies, which makes it challenging to study their structure and dynamics \citep{2013ARAA..51..511K}.
It is important to consider the types of systems that are selected for observation, as this can introduce biases, such as a focus on galaxies with AGNs, which are not representative of the overall galaxy population. These technical limitations can make it difficult to obtain detailed and accurate interpretation of the BH-bulge mass relation.

Large-scale cosmological simulations have been successful in reproducing many aspects of the Universe with a high degree of accuracy. One aspect that has been well reproduced is the large-scale structure of the Universe, including the distribution and size of galaxies, clusters of galaxies, and cosmic voids \citep{2018MNRAS.474.3976G,2018MNRAS.475..648P}. These simulations have also been successful in reproducing the observed distribution of matter in the Universe, including the distribution of dark matter, which is difficult to detect directly \citep[][and references therein]{2020NatRP...2...42V,2022LRCA....8....1A}.
In particular, we use: EAGLE \citep{2015MNRAS.446..521S,2015MNRAS.450.1937C}, Illustris \citep{2014MNRAS.445..175G,2014MNRAS.444.1518V}, TNG100, TNG300 \citep{2018MNRAS.475..676S,2018MNRAS.477.1206N,2018MNRAS.480.5113M,2018MNRAS.475..624N,2018MNRAS.473.4077P,2018MNRAS.475..648P}, Horizon-AGN \citep{2014MNRAS.444.1453D,2016MNRAS.463.3948D}, and SIMBA \citep{2019MNRAS.486.2827D}.

Our aim in this work is to setup the methodology to constrain the SMBHB properties using future PTA observations. We concentrate on the BH-bulge mass relation and test for its redshift dependence.
Existing formulations of the BH-bulge mass relation as a function of redshift for $z<5$ can be improved in light of ,e.g., the recent developments in cosmological simulations and observations from new instruments.
Thus, we formulate an equation for the BH-bulge mass relation taking into account the redshift of the system and apply this equation to fit for BH and galaxy stellar mass data from several large-scale cosmological simulations.
This BH-bulge mass relation with redshift dependence is then used in an analytical astrophysical model to compute the intensity of the GWB generated by a population of SMBHBs focusing on the PTA frequency range. Bayesian analysis is used to find the posterior of all the parameters of this GWB model. We also fix the BH-bulge mass parameters to those fitted to the cosmological simulations to constrain the posteriors of other parameters.

The paper is organized as follows. Section \ref{sec:Model} describes the astrophysical model to compute the GWB formed by the mergers of a population of SMBHBs in a parametric form using the GSMF, pair fraction and merger timescale. 
Section \ref{sec:astro} focusses on the relation between the galaxy bulge and central black hole mass, where we review the redshift independent relation and extend it by fitting to results from large-scale cosmological simulations.
In Section \ref{sec:analysis}, the analysis setup, the priors motivated by observations and large-scale cosmological simulations, and the simulation of GWB detections with different strains are described.
We present our results in Section \ref{sec:results} for the different GWB strains and also study the impact of using fixed BH-bulge mass parameters fitted to cosmological simulations. 
Finally, Section \ref{sec:Conclusions} outlines the conclusions.

\section{GWB characteristic strain}
\label{sec:Model}

For a population of SMBHBs the characteristic spectrum of the GWB was expressed in \cite{2001astro.ph..8028P} as 
\begin{equation}
	h_c^2(f) = \frac{4G}{\pi c^2 f} \int_0^\infty \dd z \int_0^\infty \dd \mathcal{M} \dv{E}{f_r} \frac{\dd[2]{n}}{\dd z \dd \mathcal{M}} \,, \label{eqn:hcf}
\end{equation}
where $f $ is the frequency, $G$ is the Newton's constant, $ c $ is the speed of light, and $z$ is the redshift. The chirp mass $\mathcal{M} $ is given as
\begin{equation}
	\mathcal{M} = \frac{(M_1M_2)^{3/5}}{(M_1+M_2)^{1/5}} \,,
\end{equation}
where $M_1,M_2$ are the individual SMBH masses in the binary system. The amount of energy emitted as GWs by each individual binary $\dv{E}{f_r}$ is dependent on the GW frequency in the source rest frame $(f_r = (1+z)f)$. The SMBHB merger rate (comoving number density in Mpc$^3$ of SMBHB mergers) per unit redshift and chirp mass $\frac{\dd[2]{n}}{\dd z \dd \mathcal{M}}$ can be derived from astrophysical observables or from a phenomenological function.

Below we summarize the parametric model from \cite{2017MNRAS.470.1738C} and \cite{2019MNRAS.488..401C}, which is extended by a parameter describing the redshift dependent evolution of the BH-bulge relation, see Section \ref{sec:astro}.

\subsection{Individual binary}

\subsubsection{Analytic model and fitting function}

Using the formalism of \cite{2017MNRAS.470.1738C} we write $\dv{E}{f_r}$ in terms of sum of harmonics at each eccentricity $e_n$ at each orbital frequency of the binary as
\begin{equation}
	\dv{E}{f_r} = \frac{\mathcal{M}^{5/3} (\pi G)^{2/3}}{3(1+z)f^{1/3}} \sum_{n=1}^\infty \frac{g_n(e_n)}{F(e_n)(n/2)^{2/3}} \,,
\end{equation}
where
\begin{eqnarray}
	F(e)   &=& \frac{1+(73/24)e^2+(37/96)e^4}{(1-e^2)^{7/2}} \,,  \\
	g_n(e) &=& \frac{n^4}{32} \Big[  \big(J_{n-2} (ne) - 2eJ_{n-1} (ne) + \frac{2}{n} J_n (ne)+ 2eJ_{n+1} (ne)  \nonumber \\ & & \quad \quad - J_{n+2} (ne) \big)^2 + \ \big(1 - e^2 \big) \big(J_{n-2} (ne) - 2J_n (ne) \nonumber \\ & & \quad \quad + J_{n+2} (ne)\big)^2 + \frac{4}{3n^2} J_n^2 (ne) \Big]
\end{eqnarray}
and $J_n$ is the first kind of $n^{th}$ Bessel function.

To increase the computational efficiency \cite{2017MNRAS.470.1738C} use the characteristic strain spectrum $h_{c,0}(f)$ of a reference SMBHB with $e_0 = 0.9$ at $f_0 = 10^{-10}$ and peak frequency $f_{p,0}$. For a generic SMBHB with $e_t$ at $f_t \neq f_0$ the strain can be computed as

\begin{equation}
	h_c(f) = h_{c,0} \Bigg( f \frac{f_{p,0}}{f_{p,t}} \Bigg)   \Bigg(\frac{f_{p,t}}{f_{p,0}} \Bigg)^{-2/3}
\end{equation}
with the peak frequency 
\begin{equation}
	f_p = \frac{1293 f}{181} \Bigg[ \frac{e^{12/19}}{1-e^2} \Bigg( 1+ \frac{121e^2}{304} \Bigg)^{870/2299} \Bigg]^{3/2} \,.
\end{equation}

A trial analytic function for the characteristic spectrum for the reference SMBHB with $\Bar{f} = f/ (10^{-8}$Hz$)$ can be written as

\begin{equation}
	h_{c,\text{fit}}(f) = a_0 \Bar{f}^{a_1} e^{-a_2\Bar{f}} + b_0 \Bar{f}^{b_1} e^{-b_2\Bar{f}}  + c_0 \Bar{f}^{c_1} e^{-c_2\Bar{f}} \,.
\end{equation}
The constants $a_0,a_1,a_2,b_0,b_1,b_2,c_0,c_1,c_2$ are determined by the fit and are given in \cite{2017MNRAS.470.1738C}. By considering SMBHBs with different redshifts and chirp masses, we get the characteristic spectrum of a population of SMBHBs as

\begin{eqnarray}
	h_c^2(f) &=& \int_0^\infty \dd z \int_0^\infty \dd \mathcal{M} \frac{\dd[2]{n}}{\dd z \dd \mathcal{M}} h_{c,\text{fit}}^2 \Big( f \frac{f_{p,0}}{f_{p,t}} \Big)   \Big(\frac{f_{p,t}}{f_{p,0}} \Big)^{-4/3} \nonumber \\ && \quad \times \Big(\frac{1+z}{1+z_0} \Big)^{-1/3} \Big(\frac{\mathcal{M}}{\mathcal{M}_0} \Big)^{5/3}
\end{eqnarray}

\subsubsection{Stellar environment}

The GWB energy spectral shape is affected by the environmental coupling. A super-efficient inspiral can cause a bend in the GWB spectrum in the PTA frequency range \citep{2013CQGra..30v4014S,2014MNRAS.442...56R,Huerta:2015pva,2017MNRAS.470.1738C}. 
At short separations the gravitational radiation starts to dominate the binary evolution, after a phase where the energy loss was driven by
interactions with stellar or gaseus environment \citep{Sampson:2015}.
We now consider that in our model stellar hardening dominates at low frequency until it is overtaken by the GW emission at the transition frequency

\begin{equation}
	f_t = 0.356 \text{ nHz  } \Bigg( \frac{1}{F(e)} \frac{\rho_{i,100}}{\sigma_{200}} \rho_0 \Bigg)^{3/10} \mathcal{M}_9^{-2/5} \,, \label{eqn:f_t}
\end{equation}
where the chirp mass $\mathcal{M}_9 = \mathcal{M}/(10^9)$ is rescaled, $\rho_{i,100}$ is the stellar density of the environment within the SMBHB influence radius, the additional multiplicative factor $\rho_0$ includes all systematic uncertainties while estimating $\rho_{i,100}$ and $\sigma_{200}$ is the stellar central velocity dispersion in the galaxy, which are given by
\begin{eqnarray}
	\rho_{i,100} &=& \frac{\rho_i}{100} \approx  \Big( \frac{2M_{\rm BH}}{M} \Big)^{\gamma/(\gamma-3)} \frac{(3-\gamma)M}{400\pi a^3 } \,, \\
	\sigma_{200} &=& \frac{\sigma}{200} = \frac{261 }{200}  \Big( \frac{M_{\rm BH}}{10^9} \Big)^{0.228} \,.
\end{eqnarray}
the stellar density distribution's inner slope is given by $\gamma \in (0.5 , 2)$, $a$ is the characteristic radius and $M$ is the total bulge mass of the galaxy, which are expressed as 
\begin{eqnarray}
	a &=& 239 (2^{1/(3-\gamma)}-1) \Big( \frac{M}{10^9} \Big)^{0.596} \,, \\
	M &=& 1.84 \times 10^{11}  \Big( \frac{M_{\rm BH}}{10^9} \Big)^{0.862} \,.
\end{eqnarray}

\subsection{Merger rate}

The merger rate in Eqn (\ref{eqn:hcf}) can be written in terms of SMBHB mass as
\begin{equation}
	h_c^2(f) = \frac{4G}{\pi c^2 f} \int_0^\infty \dd z \int_0^\infty \dd{M_{\rm BH}} \dv{E}{f_r} \int_0^1   \frac{\dd[3]{n}}{\dd z \dd{M_{\rm BH}} \dd q_{\rm BH}} \dd q_{\rm BH} \,, \label{eqn:hcf-mod}
\end{equation}
where SMBHB merger rate and black hole mass are
\begin{eqnarray}
	\frac{\dd[3]{n}}{\dd z \dd{M_{\rm BH}} \dd q_{\rm BH}}   &=& \frac{\dd[3]{n_G}}{\dd z \dd{M} \dd q} \frac{\dd M}{\dd M_{\rm BH}} \frac{ \dd q}{ \dd q_{\rm BH} }  ,\label{eqn:BH_mergerrate} \\
	M_{\rm BH}  &=& \frac{ \mathcal{M} (1+q_{\rm BH})^{1/5} }{ q_{\rm BH}^{3/5} } \,.
\end{eqnarray}

$M_{\rm BH}$ can be parameterized using galaxy bulge mass as shown in Section \ref{sec:astro}.
An astrophysical observable based description of the galaxy merger rate is given in \cite{2013CQGra..30v4014S,2016MNRAS.463L...6S,2019MNRAS.488..401C} as 

\begin{equation}
	\frac{\dd[3] n_G}{\dd z' \dd M \dd q} = \frac{\Phi(M,z) \ \  \mathcal{F}(z,M,q) }{M \ \ln(10) \ \  \tau (z,M,q)} \frac{\dd t_r}{\dd z} \,, \label{eqn:d3ng}
\end{equation}

where $q$ is the galaxy binary mass ratio with the primary galaxy mass $M$, $\Phi(M,z) = (\dd n_G /\dd \log M)_z $ is the GSMF estimated at redshift $z$, the differential pair fraction of the galaxy binaries is $\mathcal{F}(z,M,q) = (\dd f/ \dd q)_{z,M}$ and the merger timescale $\tau (z,M,q) = \int_{z'}^z (\dd t/ \dd \Tilde{z}) \dd \Tilde{z}$ is obtained by integrating over the instantaneous redshift $\dd \Tilde{z}$ between the redshifts at the start $z'$ and end $z$ of the galaxy merger. Using a flat $\Lambda \text{CDM}$ model one finds
\begin{equation}
	\frac{\dd t}{\dd \Tilde{z}} = \frac{1}{ H_0 (1+\Tilde{z}) \sqrt{ \Omega_M (1+\Tilde{z})^3 + \Omega_k (1+\Tilde{z})^2 + \Omega_\Lambda}} \,.
\end{equation}

Here we use energy density ratios $\Omega_M =0.3, \Omega_k =0, \Omega_\Lambda=0.7$ and Hubble constant $H_0=70$ kmMpc$^{-1}$s$^{-1}$.

\subsubsection{Galaxy Stellar Mass Function}

The Galaxy Stellar Mass Function (GSMF) describes the number density of galaxies as a function of their stellar mass. The assembly of stellar mass and the evolution of the stellar formation rate through cosmic time can be traced using the GSMF and is a major estimate of the characteristics of the galaxy population.

This astrophysical observable can be parameterized and fitted in the form of a Schechter function \citep{2016ApJ...830...83C}. To take into account redshift evolution we can write the GSMF using parameters from \cite{2015MNRAS.447....2M} as 
\begin{equation}
	\Phi(M,z) = \ln (10) 10^{\Phi_0+z\Phi_I} \Big( \frac{M}{M_0} \Big)^{1+\alpha_0+ z \alpha_I }  \exp \Big( \frac{-M}{M_0} \Big) \,.
\end{equation}

\subsubsection{Pair fraction}

The differential pair fraction of the galaxy binaries at $M$ and $z$ with respect to $q$ can be written as \citep{2017MNRAS.470.3507M}
\begin{equation}
	\mathcal{F}(z,M,q) = f'_0 \Big( \frac{M}{10^{11} } \Big)^{\alpha_f} (1+z)^{\beta_f} q^{\gamma_f} = \frac{\dd f_{\text{pair}}(z,M)}{\dd q} 
\end{equation}
with $f_0 = f'_0 \int q^{\gamma_f} \dd q$. Integrating over $q$ then gives
\begin{equation}
    f_{\text{pair}} = f_0 \Big( \frac{M}{10^{11} } \Big)^{\alpha_f} (1+z)^{\beta_f} \,.
\end{equation}

\subsubsection{Merger timescale}

The timescale of the evolution of a binary galaxy from the dynamical friction can be used to approximate the full merger timescale, which can be written using a parameterisation with $\tau_0, \alpha_\tau, \beta_\tau, \gamma_\tau$ as
\begin{eqnarray}
	\tau(z,M,q) = \tau_0 \Bigg( \frac{M \ h_0}{0.4 \times 10^{11} } \Bigg)^{\alpha_\tau} 	(1+z)^{\beta_\tau} q^{\gamma_\tau}
\end{eqnarray}
where $h_0 = 0.7$ is the Hubble parameter.

Substituting these observables into Eqn (\ref{eqn:d3ng}) gives
\begin{eqnarray}
	\frac{\dd[3] n_G}{\dd z' \dd M \dd q} &=& \frac{10^{\Phi_0+z\Phi_I} f'_0}{M_0\tau_0} \Bigg( \frac{0.4}{h_0} \Bigg)^{\alpha_\tau} 
	\Bigg( \frac{M}{10^{11}}  \Bigg)^{ \alpha_f - \alpha_\tau} 
	\Bigg( \frac{M}{M_0} \Bigg)^{\alpha_0+ z \alpha_I} \nonumber \\ & & \quad \times
	e^{-M/M_0} (1+z)^{\beta_f - \beta_\tau} q^{\gamma_f- \gamma_\tau} \frac{\dd t}{\dd z} \,.
\end{eqnarray}


\section{Astrophysics of SMBH Mass}
\label{sec:astro}

The final ingredient to describe the SMBHB merger rate in Eqn (\ref{eqn:BH_mergerrate}) is a relation between the galaxy stellar mass and the central black hole mass. We first express the bulge mass of a galaxy using its total stellar mass, and then use the resulting BH - bulge mass relation to extract the BH mass needed for the computation of the merger rate.

The fraction of the total stellar mass assigned to the bulge mass depends on the galaxy morphology and galaxy mass regime.
For this work the phenomenological stellar-bulge mass relation (\cite{2014MNRAS.443..874B,2016MNRAS.463L...6S}) is used
\begin{equation}
	M_{\rm bulge} =  \left\{
	\begin{array}{l  l}
		\Big(\frac{\sqrt{6.9}}{(\log M - 10)^{1.5}} \exp{ \frac{-3.45}{\log M - 10} } + 0.615\Big)M & \text{if} \log M > 10 \\
		0.615 \ M & \text{if} \log M \leq 10. \\
	\end{array} 
	\right. \label{eqn:stellar-bulge}
\end{equation}
This relation focusses on spherical and elliptical galaxies, which dominate the PTA GWB signal. Higher mass galaxies $M \leq 10^{10} M_\odot$ have been observed to be correlated with the size of the bulge and disk, while lower mass galaxies do not.

\subsection{Large-scale cosmological simulations}

In this paper we investigate the differences in the BH-bulge mass relations produced in EAGLE, Illustris, TNG100, Horizon-AGN, SIMBA, and TNG300, and quantify the evolution of the relation with redshift.
The galaxy stellar mass and the corresponding SMBH mass from the simulations are given in \cite{2021MNRAS.503.1940H}.
The conversion of the stellar mass of the galaxies into their bulge mass is done using Eqn (\ref{eqn:stellar-bulge}). Thus, the BH-bulge mass relation is connected to the BH-galaxy stellar mass relation.

Cosmological simulations model the dark matter and baryonic contents of the Universe in an expanding space-time.
All the simulations studied in this paper have a volume of $\geqslant 100^{3}\, \rm cMpc^{3}$, a dark matter mass resolution of $\sim 5\times 10^{6}-8\times 10^{7}\, \rm M_{\odot}$, and a spatial resolution of $1-2$ ckpc. As such, the simulations capture the time evolution of the galaxies with a total stellar mass in the range $M=10^{9}-10^{11-12}\, \rm M_{\odot}$ and their BHs. Baryonic processes taking place at small scales below the galactic scale are modelled as subgrid physics (e.g., supernova (SN) and AGN feedback). Although theoretically based on the same idea, these processes are modelled differently in each simulation. For example, AGN feedback releases energy in the BH surrounding but the implementation in the simulation can rely on the injection of thermal energy only, thermal and kinetic energy or momentum in a given direction to mimic an outflow or jet. 
The subgrid physics of the simulations impact the evolution of both galaxies and BHs \citep{2021MNRAS.503.1940H,2022MNRAS.509.3015H}. 

There is no consensus on the shape nor on the time evolution of the BH-bulge relation produced by the EAGLE, Illustris, TNG100, Horizon-AGN, SIMBA, and TNG300 simulations \citep{2021MNRAS.503.1940H,2022MNRAS.511.3751H}. 
The shape of the BH-bulge relation in the low-mass end ($M\leqslant 10^{10.5}\, \rm M_{\odot}$) is mainly driven by BH seeding mass, strength of SN feedback and BH accretion modelling. The massive end is affected by the modelling of AGN feedback and BH accretion.
Half of the simulations have more massive BHs at high redshift than at $z=0$ at fixed galaxy stellar mass. The other simulations follow the opposite trend.
On average, the time evolution of the relation depends on whether BHs grow more efficiently than their host galaxies (see summary in Fig.~11 in \cite{2021MNRAS.503.1940H}).

\subsection{Empirical BH--bulge mass relation}

The BH-bulge mass relation is a key quantity for our understanding of the co-evolution of galaxies and their central black holes. The redshift independent BH-bulge mass relation \citep{2013ARAA..51..511K} which is usually used in the literature is given by

\begin{eqnarray}
	M_{\rm BH} &\sim& \mathcal{N} \Bigg\{  \Big( \frac{M_{\rm bulge}}{10^{11}{\rm M}_\odot} \Big)^{\alpha} 10^\beta , \varepsilon \Bigg\} \\
	\log_{10}M_{\rm BH} &=& \alpha \log_{10} \Big( \frac{M_{\rm bulge}}{10^{11}{\rm M}_\odot} \Big) + \beta \,,
\end{eqnarray}
where $M_{\rm BH}$ is the mass of the SMBH at the centre of the galaxy with bulge mass $M_{\rm bulge}$ and $\mathcal{N}$ denotes a Gaussian distribution.
$\alpha$ and $\beta$ are the BH-bulge mass parameters that determine the slope and normalization of the relation respectively. On the logarithmic scale the relation becomes a straight line with scattering $\varepsilon$ where the parameters can be deduced from a least squares fit. Reviews on different models and parameters of this relation can be found in \cite{Sesana:2013} and \cite{2019ApJ...887..245S}.

\subsection{BH--bulge mass relation with redshift dependence}

Our goal is to formulate a parametric redshift dependent BH-bulge mass relation for $z \leq 5 $ since the GWB should be detectable with PTA in this redshift range.
We propose a relation given by
\begin{eqnarray}
	M_{\rm BH} &\sim&  \mathcal{N} \Bigg\{  \Bigg( \frac{M_{\rm bulge}}{10^{11}{\rm M}_\odot} \Bigg)^{\alpha_*} 10^{\beta_*+\gamma_* z} , \varepsilon \Bigg\}  \\
	\log_{10}M_{\rm BH} &=& \alpha_* \log_{10}\Big(\frac{M_{\rm bulge}}{10^{11}{\rm M}_\odot}\Big) + \beta_* +\gamma_*z \,, \label{myeqnlog}
\end{eqnarray} 
where we consider an additional BH-bulge mass parameter $\gamma_*$, which determines the extent of the evolution of the black hole mass with redshift. Positive $\gamma_*$ values result in larger black hole masses as the redshift increases, while negative $\gamma_*$ values have the opposite effect.

This relation is based on the assumption that the BH-bulge mass relation evolves only through the normalization parameter $\beta_*$, while the slope $\alpha_*$ remains constant with redshift.
This assumption is based on the observation that the correlation between the mass of the black hole and the mass of the bulge is largely set by the processes that lead to the formation of the bulge, which happen early in the galaxy's history. These processes are not expected to change significantly over cosmic time and so the slope parameter is expected to remain relatively constant.
However, the normalization parameter is expected to change with redshift because the growth of the black hole and the bulge are linked through complex feedback processes. These feedback processes are expected to change over time as the galaxy evolves, and so the normalization parameter is expected to evolve with redshift \citep{2013ARAA..51..511K}.

\begin{figure*}
	\centering
	\vspace{1cm}
	\includegraphics[width=\textwidth]{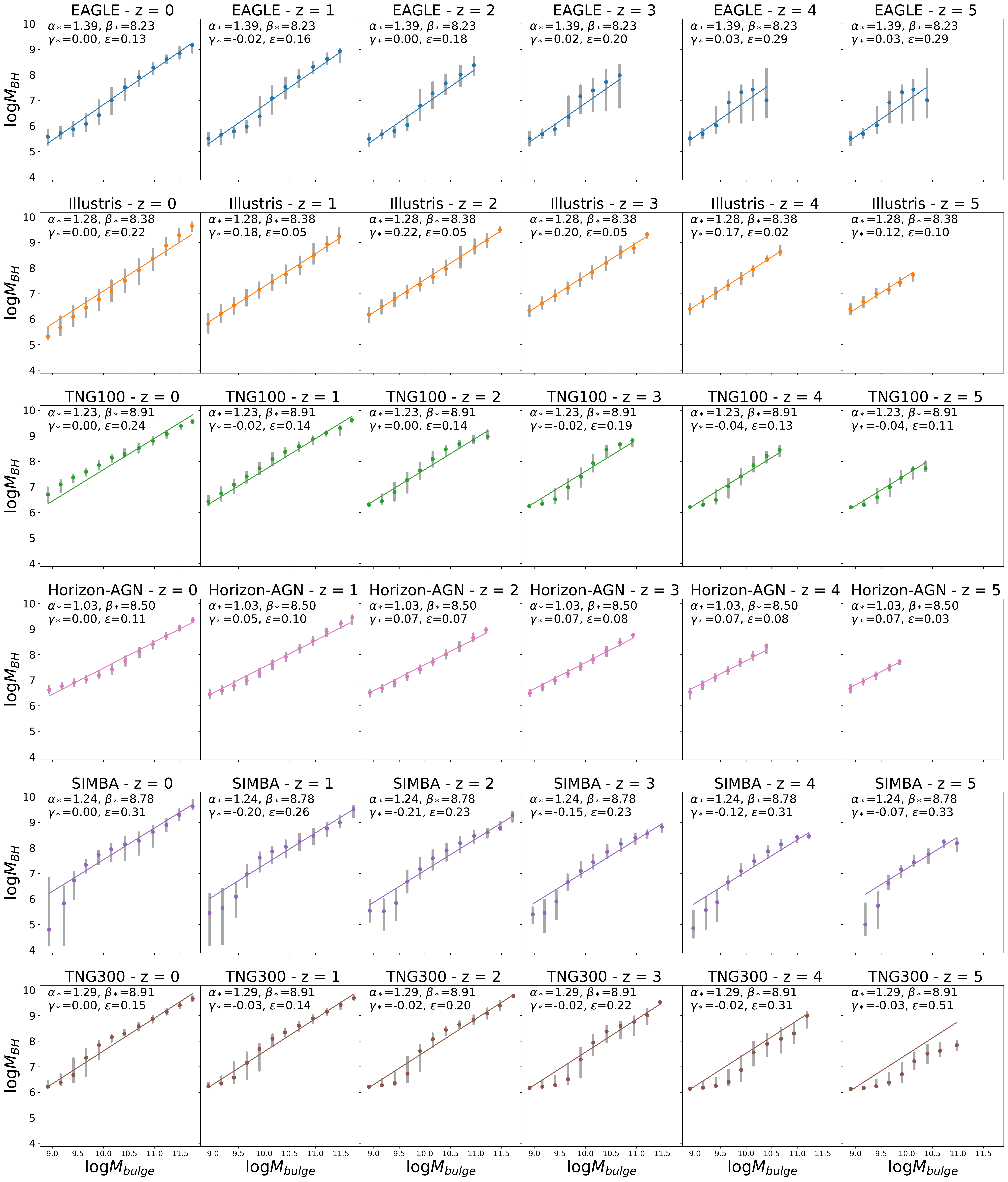}
	\caption{Best fit BH-bulge relations for the EAGLE, Illustris, TNG100, Horizon-AGN, SIMBA, and TNG300 simulations as they evolve with redshift. The BH and bulge masses with uncertainties from the simulations are consistent with the BH-bulge relations at different redshifts.}
	\label{fig:z_all_sim}
	\vspace{1cm}
\end{figure*}

\begin{figure*}
	\centering
	\includegraphics[width=\textwidth]{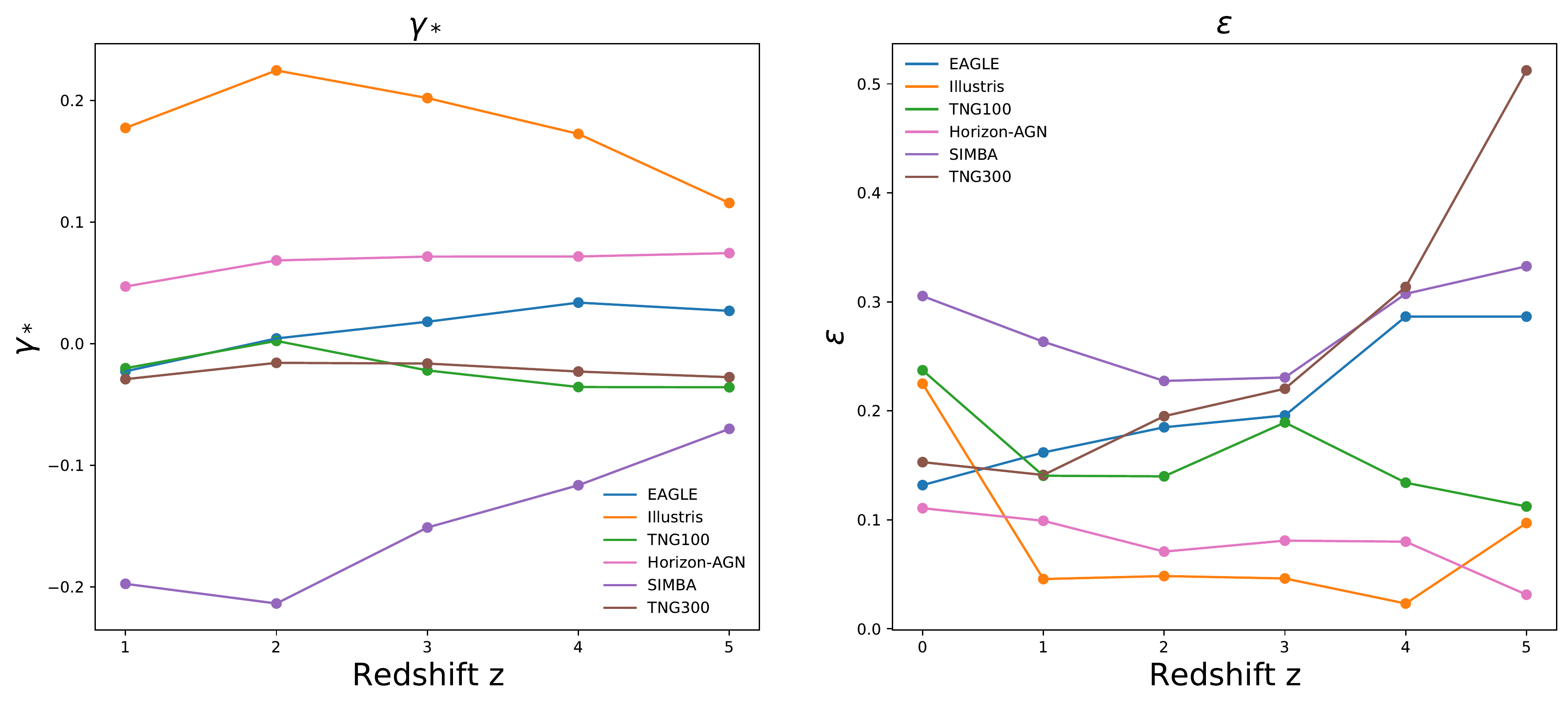}
	\caption{Variation of best fit $\gamma_*$ and $\varepsilon$ values with redshift for the EAGLE, Illustris, TNG100, Horizon-AGN, SIMBA, and TNG300 simulations. The values are approximately constant across the redshift range for a given simulation, thus, one set of parameters $(\alpha_*,\beta_*,\gamma_*,\varepsilon)$ (see \autoref{tab:sims}) can be used to represent the corresponding simulation.}
	\label{fig:para_ga_ep}
\end{figure*}

Eqn (\ref{myeqnlog}) is fitted to the SMBH and galactic bulge masses for each of the six cosmological simulations separately. For a given simulation $\alpha_*$ is the slope in the logarithmic scale given by the linear least squares fit over all redshifts $z \leq 5$. $\beta_*$ is the intercept at $M_{\rm bulge} = 10^{11} M_\odot$ of the least squares fit at $z=0$. The intercepts at different redshifts is used to compute $\gamma_ *$. The amount of scattering of the SMBH mass from the phenomenological fit in Eqn (\ref{myeqnlog}) is denoted by $\varepsilon$.  \autoref{tab:sims} lists the BH-bulge mass relation parameters for these cosmological simulations.
The variation of the masses for these simulations are plotted in \autoref{fig:z_all_sim} as they evolve with redshift. \autoref{fig:para_ga_ep} shows the best fit values of $\gamma_*$ and $\varepsilon$ for the simulations at different redshifts. The values are approximately constant across the different redshifts, thus allowing us to use the average as a set of parameters that can approximately reproduce the masses from the simulations at all redshifts.

An alternative redshift dependent BH-bulge mass relation used by e.g., \cite{2016ApJ...816...37V} is written as
\begin{eqnarray}
	M_{\rm BH} &\sim&  \mathcal{N} \Bigg\{  \Bigg( \frac{M_{\rm bulge}}{10^{11}{\rm M}_\odot} \Bigg)^{\alpha_*} 10^{\beta_*} (1+z)^{\gamma_*} , \varepsilon \Bigg\}  \\
	\log_{10}M_{\rm BH} &=& \alpha_* \log_{10}\Big(\frac{M_{\rm bulge}}{10^{11}{\rm M}_\odot}\Big) + \beta_* +\gamma_* \log_{10} (1+z) \,.
\end{eqnarray}

Fitting the black hole and bulge masses to this relation, we obtain large variability of the parameters and higher scattering values for most of the simulations we have used. Thus, in this work we use the previous relation that produces stable parameter values at different redshifts and lower scattering values consistent with the simulations.

We note that our relation (Eqn (\ref{myeqnlog})) is a simple approximation to the simulations extending the redshift independent BH-bulge relation. Therefore, our best fit values may not be fully representative of the results from the simulations. This caveat should be kept in mind with the results presented in Section \ref{sec:results}.

\begin{table}
	\centering
	
	\begin{tabular}{|l| c| c| c| c|} 
		\hline \footnotesize
		Simulation &  $\alpha_*$ & $\beta_*$ & $\gamma_*$ & $\varepsilon$ \\ [0.5ex] 
		 \hline
		EAGLE      & $1.39 \pm 0.027$ & $8.23 \pm 0.039$ & $ 0.01^{ + 0.022}_{ - 0.035}$ & $0.21^{ + 0.079}_{ - 0.076}$ \\ [1ex]
		Illustris  & $1.28 \pm 0.040$ & $8.38 \pm 0.088$ & $ 0.18^{ + 0.046}_{ - 0.063}$ & $0.08^{ + 0.144}_{ - 0.058}$ \\ [1ex]
		TNG100     & $1.23 \pm 0.022$ & $8.91 \pm 0.074$ & $-0.02^{ + 0.025}_{ - 0.014}$ & $0.16^{ + 0.078}_{ - 0.047}$ \\ [1ex]
		HorizonAGN & $1.03 \pm 0.026$ & $8.50 \pm 0.036$ & $ 0.07^{ + 0.008}_{ - 0.020}$ & $0.08^{ + 0.032}_{ - 0.048}$ \\ [1ex]
		SIMBA      & $1.24 \pm 0.046$ & $8.78 \pm 0.063$ & $-0.15^{ + 0.080}_{ - 0.064}$ & $0.28^{ + 0.055}_{ - 0.050}$ \\ [1ex]
		TNG300     & $1.29 \pm 0.019$ & $8.91 \pm 0.050$ & $-0.02^{ + 0.007}_{ - 0.007}$ & $0.26^{ + 0.256}_{ - 0.115}$ \\ [1ex]
		\hline
	\end{tabular}
	\caption{Best fit parameters with uncertainties for the redshift dependent BH-bulge relation from Eqn (\ref{myeqnlog}) using BH and bulge masses from the six large-scale cosmological simulations.} \label{tab:sims}
\end{table}

\begin{figure}
	\centering
	\includegraphics[width=\linewidth]{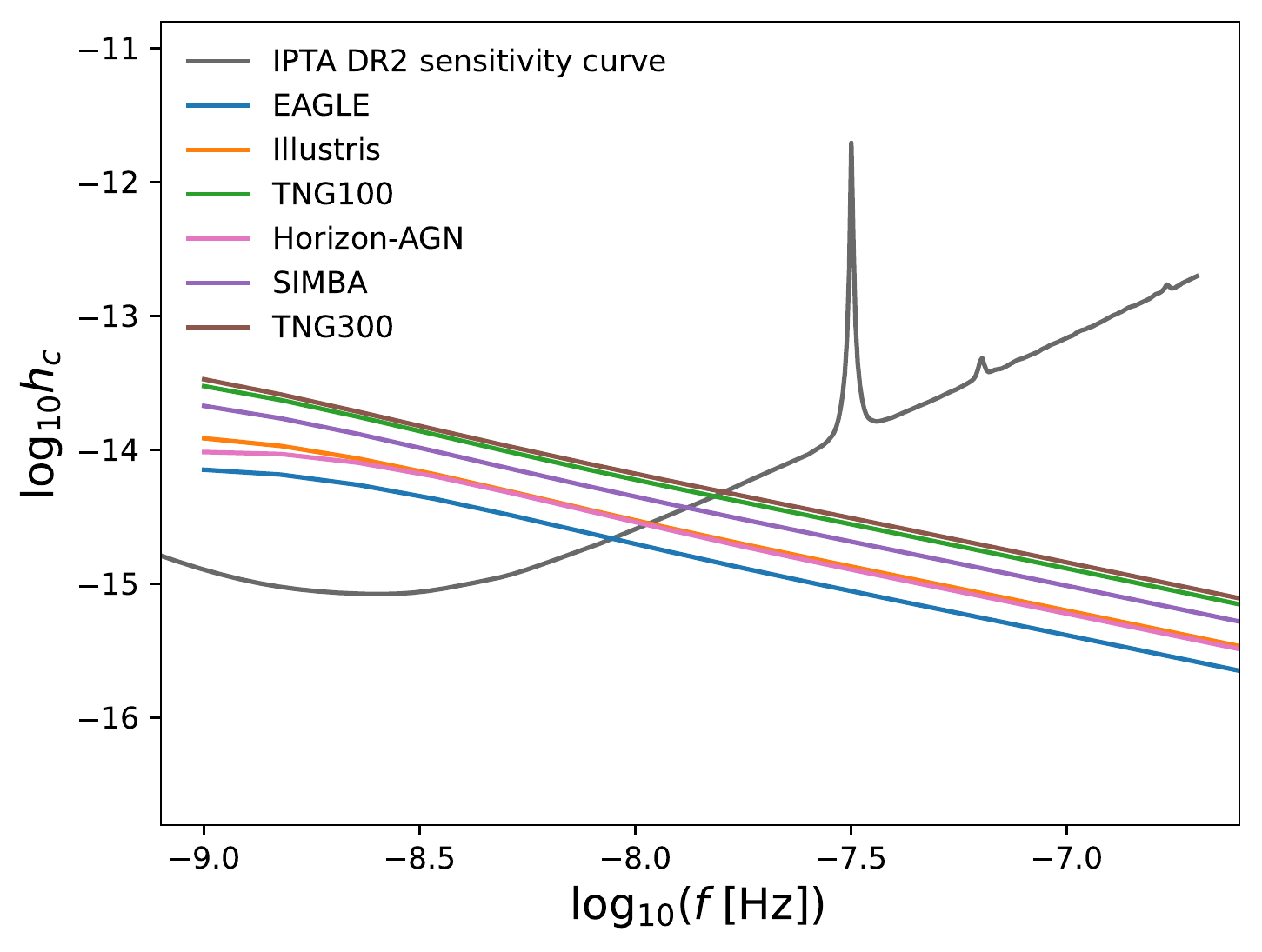}
	\caption{GWB characteristic strain spectra in the PTA range from a set of fiducial values (see \autoref{fig:gwb_para_variation}) showing the differences when using the BH-bulge mass parameters for the EAGLE, Illustris, TNG100, Horizon-AGN, SIMBA, and TNG300 simulations respectively. An analytic sensitivity curve from the IPTA DR2 \citep{2022MNRAS.510.4873A} is plotted to guide the eye.}
	\label{fig:gwb_sim}
\end{figure}

\begin{figure}
	\centering
	\includegraphics[width=\linewidth]{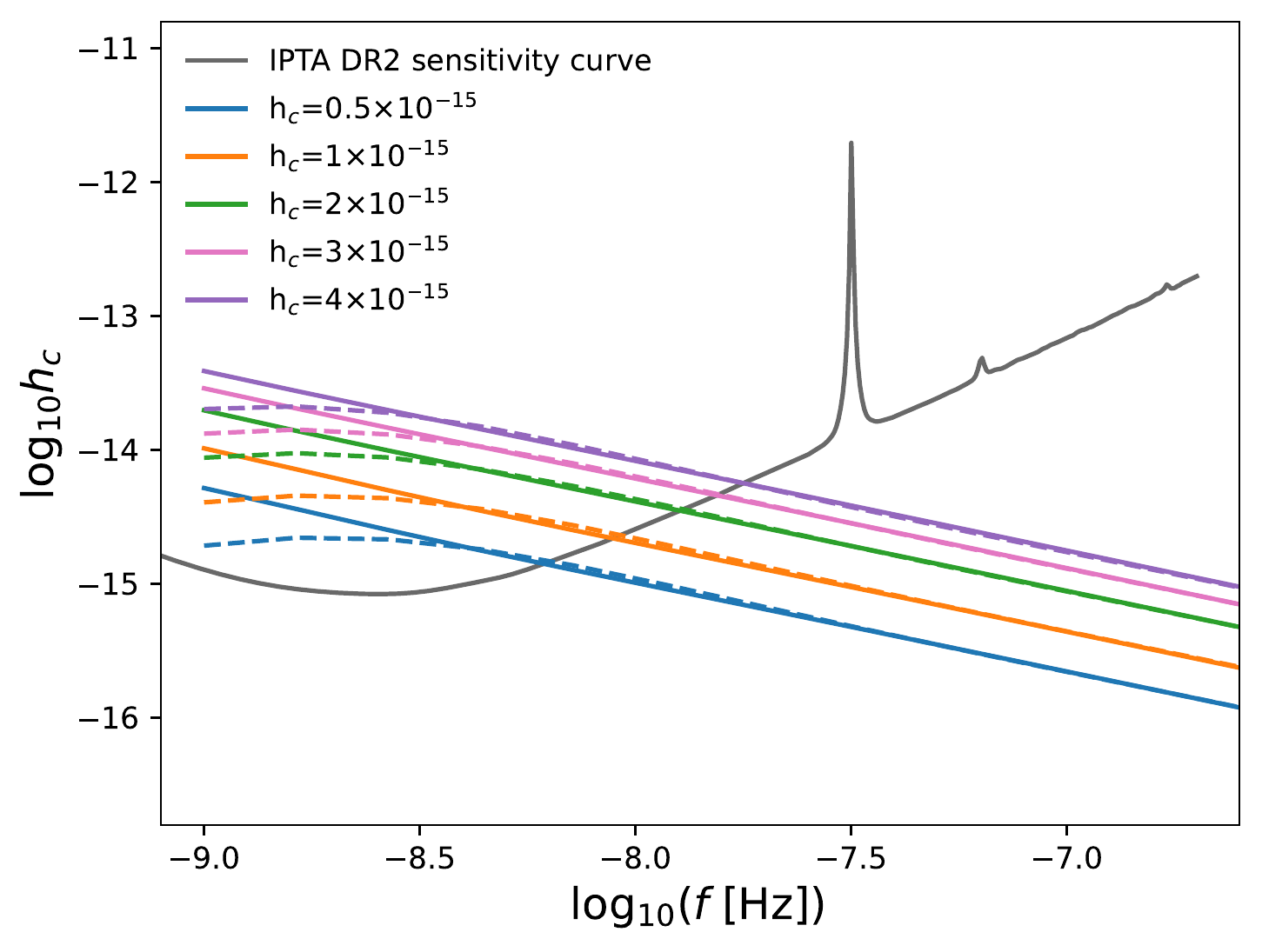}
	\caption{Simulated GWB detections with different characteristic spectra in the PTA range used in the Bayesian analysis. \autoref{tab:parameter_val} provides a non-unique set of parameter values for these spectra. To guide the eye the analytic sensitivity curve from the IPTA DR2 \citep{2022MNRAS.510.4873A}.}
	\label{fig:gwb_sim_input}
\end{figure}

\section{Bayesian analysis setup}
\label{sec:analysis}

Using all the parts described above the characteristic spectrum can be computed as a function (Eqn (\ref{eqn:hcf-mod})) with 19 parameters which can be estimated from astrophysical observables. These parameters are five GSMF parameters $\Phi_0, \Phi_I, M_0, \alpha_0,\alpha_I$, four pair fraction parameters $f_0,\alpha_f,\beta_f,\gamma_f$, four merger timescale parameters $\tau_0, \alpha_\tau,\beta_\tau,\gamma_\tau$, four BH-bulge mass parameters $\alpha_*,\beta_*,\gamma_*, \varepsilon$, and two parameters $e_0,\rho_0$ related to the individual binary GW emission.

In order to find the redshift volume that PTA can probe for galaxy and SMBH mergers we consider $z_m$, which is the maximum redshift that is used to compute the volume, as an additional parameter to see the change in GWB characteristic strain if the volume is larger or smaller.

The effect of each of the 20 astrophysical parameters on the GWB is shown in  \autoref{fig:gwb_para_variation} for a fiducial choice of values. Using the corresponding values from \autoref{tab:sims} for the six large-scale simulations the differences in the GWB spectra can be seen in \autoref{fig:gwb_sim}.

With this parametric model in hand we can set up the Bayesian analysis to use simulated PTA detections to infer what posterior constraints can be achieved for each parameter.

\subsection{Simulated GWB detections}

Different values for the 20 parameters within the prior ranges give different GWB characteristic strain. Depending on the values of the 20 parameters, we can simulate a straight line or a curve bending down at low frequency for the GWB characteristic strain in the frequency range of $10^{-9}-10^{-6}$. In our model the straight line and curve spectra are associated with circular and eccentric SMBHB populations respectively. We created datasets for these two different shapes of spectrum for strain values of $0.5\,, 1\,, 2\,, 3\,, 4 \times 10^{-15}$ at the reference frequency of $f=1/1$year ($f\approx10^{-7.5}$ Hz) as shown in  \autoref{fig:gwb_sim_input}. PTAs typically search at frequencies that are multiples of $1/T_{\rm span}$, where $T_{\rm span}$ is the total observation time span of the PTA dataset. For simplicity and computational efficiency, we use the five lowest bins with $T_{\rm span} = 25$ years. The values of the parameters used to create the different simulated spectra are chosen by hand and given in \autoref{tab:parameter_val}. These sets of parameters are non-unique and thus not necessarily representative for the given GWB spectrum.

\subsection{Likelihood function}

To simulate a detection of the GWB we assume at each frequency a Gaussian distribution of central logarithmic amplitude $\log_{10} A_{\text{det}}(f)$ and width $\sigma_{\text{det}}(f)$, which are the detection measurement errors. With the GWB computed from a trial parameter set $\log_{10} A_{\rm trial}(f)$ the likelihood function following \cite{2017MNRAS.468..404C} and \cite{2018NatCo...9..573M} can be written as
\begin{equation}
	p_{\text{det}} (d \ | A_{\text{trial}}(f)) \propto  \exp \Bigg\{ \frac{- \big(  \log_{10} A_{\text{trial}}(f) - \log_{10}  A_{\text{det}}(f)\big)^2}{2 \sigma_{\text{det}}(f)^2}  \Bigg\}.
\end{equation}

The Parallel Tempering Markov Chain Monte-Carlo (PTMCMC) Sampler \citep{justin_ellis_2017_1037579} is used with $\log_{10} A_{\text{det}}(f)$ taken from the simulated GWB datasets and $\sigma_{\rm det} = 0.09$ for the Bayesian analysis.

\begin{table}
	\centering
	\begin{tabular}{|l | c | c|} 
		\hline
		Description & Parameter  & Range \\ [0.5ex] 
		\hline
		BH-Bulge mass relation average slope & $\alpha_*$ & [1,1.5] \\
		BH-Bulge mass relation norm at $z=0$ & $\beta_*$ & [8,9] \\ 	
        BH-Bulge mass relation norm redshift evolution  & $\gamma_*$ & [-0.5,0.5] \\ 	
		BH-Bulge mass relation scatter  & $\varepsilon$ & [0.05,0.5] \\
		Maximmum redshift  & $z_m$ & [0.1,5] \\[1ex]
		\hline
	\end{tabular}
	\caption{Prior choice for the parameters of the redshift dependent BH - bulge mass relation.} \label{tab:prior}
\end{table}

\begin{table}
	\centering
	\begin{tabular}{|l | c | l|} 
		\hline
		Description & Parameter  & Range \\ [0.5ex] 
		\hline
		GSMF norm & $\Phi_0$ & [-3.4,-2.4] \\
		GSMF norm redshift evolution & $\Phi_I$ & [-0.6,0.2] \\ 	
		GSMF scaling mass  & $\log_{10} M_0$ & [11,11.5] \\ 
		GSMF mass slope  & $\alpha_0$ & [-1.5,-1] \\
		GSMF mass slope redshift evolution  & $\alpha_I$ & [-0.2,0.2] \\
		pair fraction norm  & $f_0$ & [0.01,0.05] \\
		pair fraction mass slope  & $\alpha_f$ & [-0.5,0.5] \\
		pair fraction redshift slope  & $\beta_f$ & [0,2] \\
		pair fraction mass ratio slope  & $\gamma_f$ & [-0.2,0.2] \\
		merger time norm  & $\tau_0$ & [0.1,10.0] \\
		merger time mass slope  & $\alpha_\tau$ & [-0.5,0.5] \\
		merger time redshift slope  & $\beta_\tau$ & [-3,1] \\
		merger time mass ratio slope  & $\gamma_\tau$ & [-0.2,0.2] \\
		binary eccentricity  & $e_0$ & [0.01,0.99] \\
		stellar density factor  & $\log_{10}\rho_0$ & [-2,2] \\[1ex]
		\hline
	\end{tabular}
	\caption{Prior choice for the parameters of the other astrophysical observables.}\label{tab:prior_other}
\end{table}

\subsection{Prior choice}
\label{sec:prior}

\begin{figure}
	\centering
	\includegraphics[width=\linewidth]{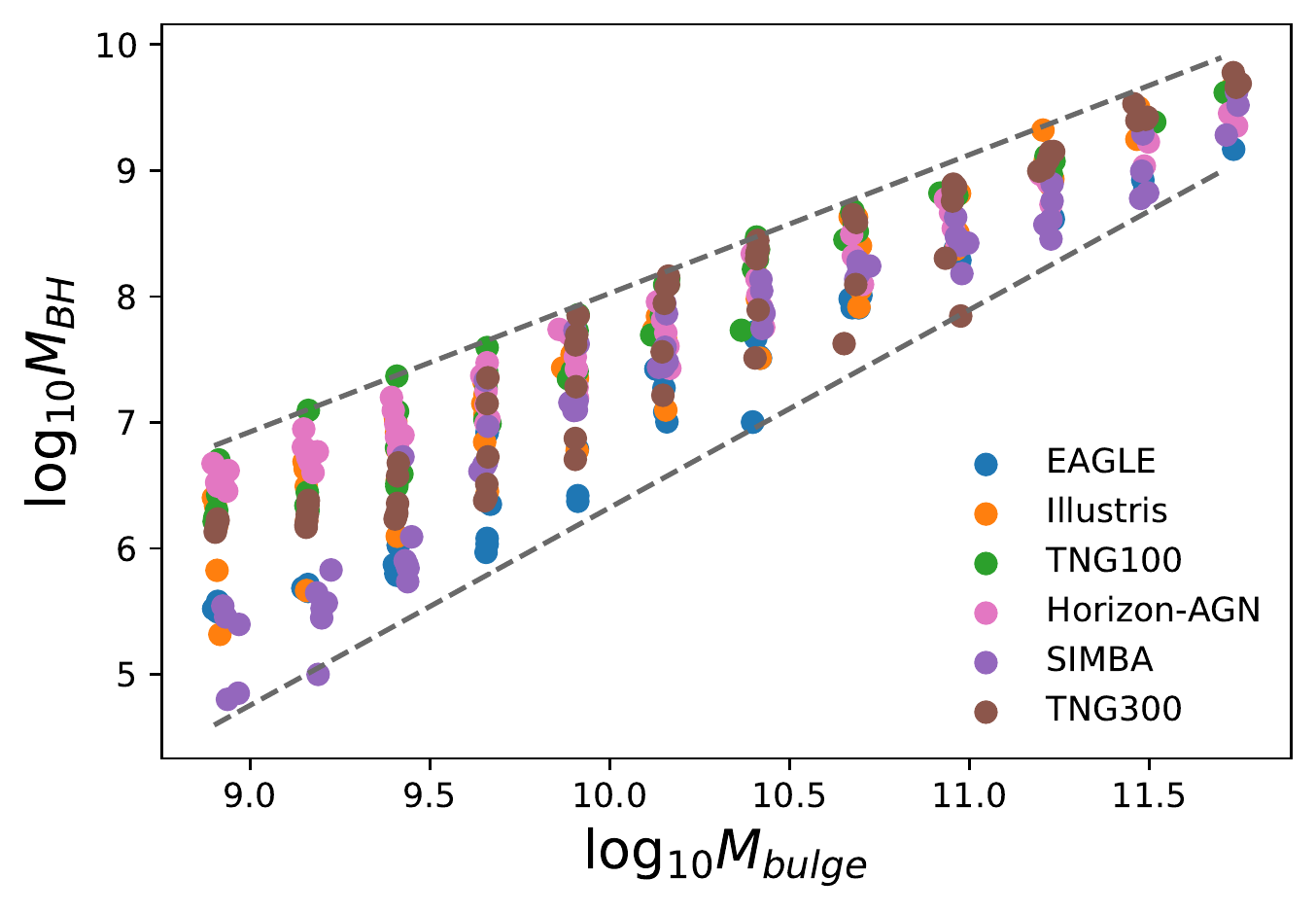}
    \caption{The BH masses of the EAGLE, Illustris, TNG100, Horizon-AGN, SIMBA, and TNG300 simulations for all redshift $z\leq 5$ as a function of the galaxy bulge mass. The maximum and minimum BH mass at the corresponding bulge mass are used to construct an allowed range for the BH-bulge relation, which is shown by the dashed lines.}
	\label{fig:prior}
\end{figure}

The prior for the BH-Bulge mass relation is constrained using all possible masses from the six different simulations for $z\leq 5$ to set the allowed range as shown in \autoref{fig:prior} and the initial test values are given the \autoref{tab:prior}. Only combinations of $\alpha_*,\beta_*$,$\gamma_*$ and $z=(0,z_{max})$ that give relations within the boundaries are accepted. This ensures that the BH-bulge relations are compatible with those from the simulations for all redshifts between 0 and $z_{max}$.

It is assumed that the redshift volume that PTAs are sensitive to is between 1.5 and 2.5. To study the effects of evolution with redshift and to test this assumption, we consider an extended range of $z_m \in [0.1,5]$, which includes the above range.

For the other parameters we adopt the same prior choice as \cite{2019MNRAS.488..401C} shown in \autoref{tab:prior_other}.

\section{Results}
\label{sec:results}

\subsection{Consistency of the cosmological simulations with GWB detections}

\begin{figure*}
	\centering
	\includegraphics[width=0.47\textwidth]{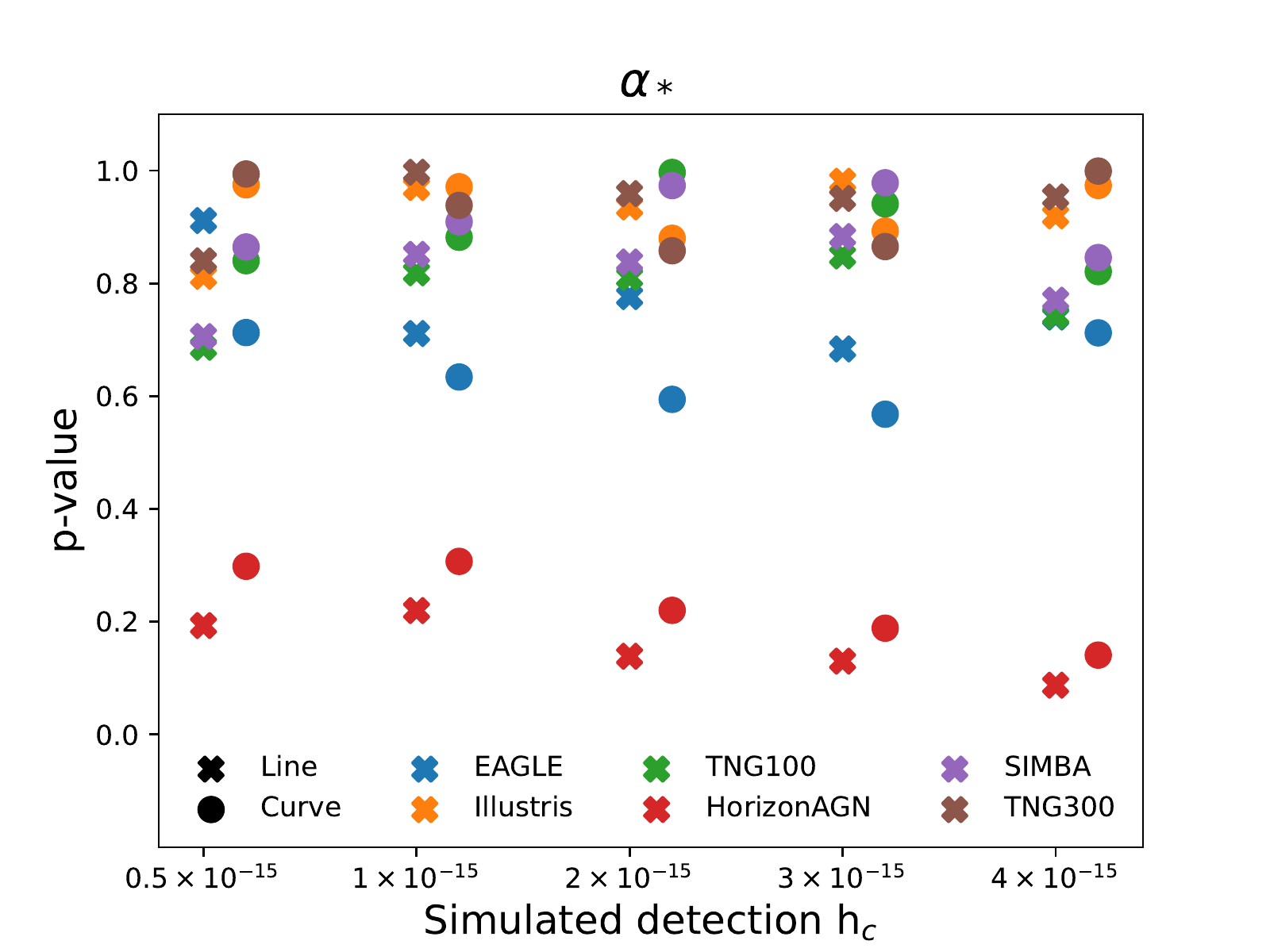}
    \includegraphics[width=0.47\textwidth]{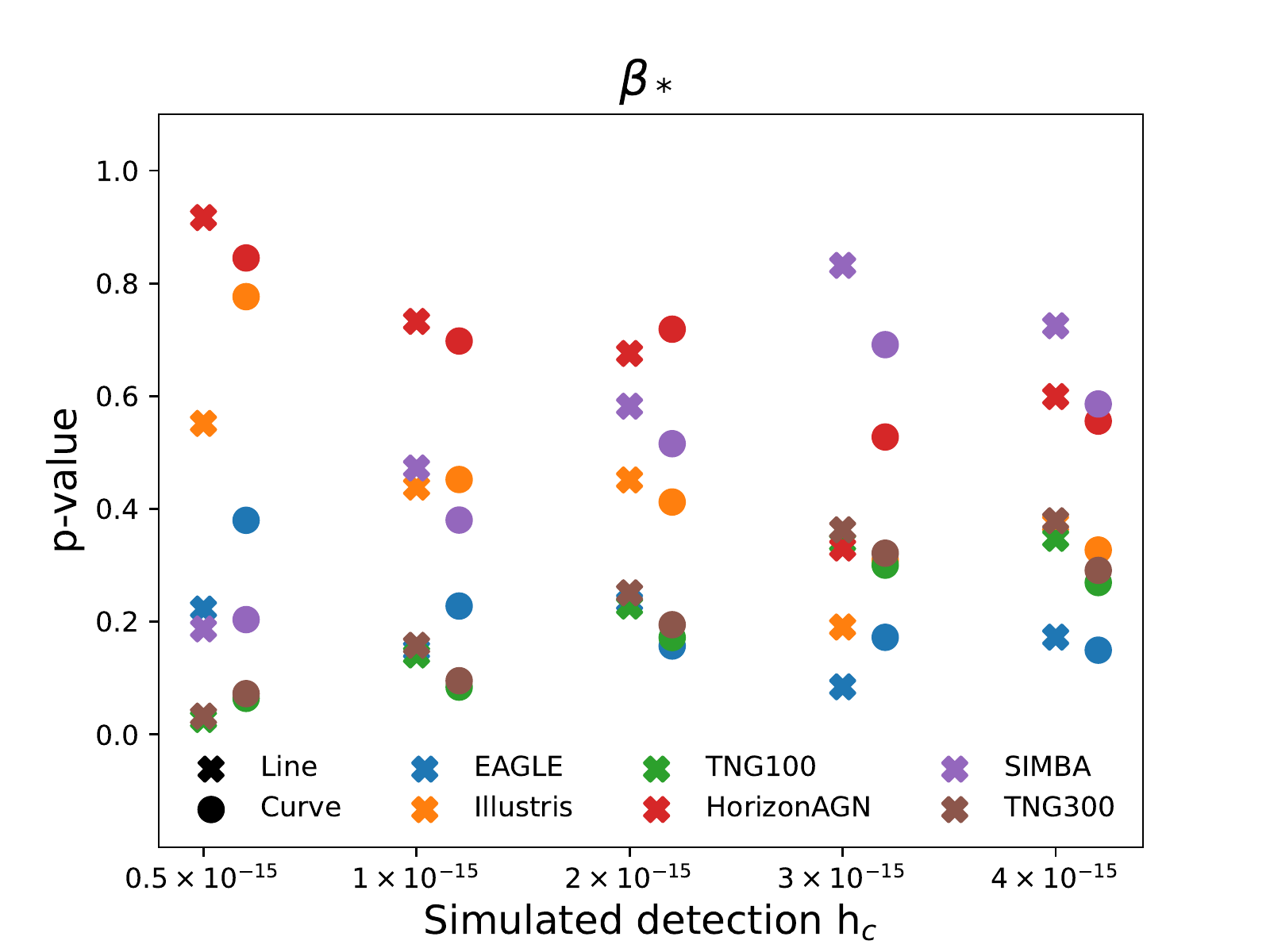}
    \includegraphics[width=0.47\textwidth]{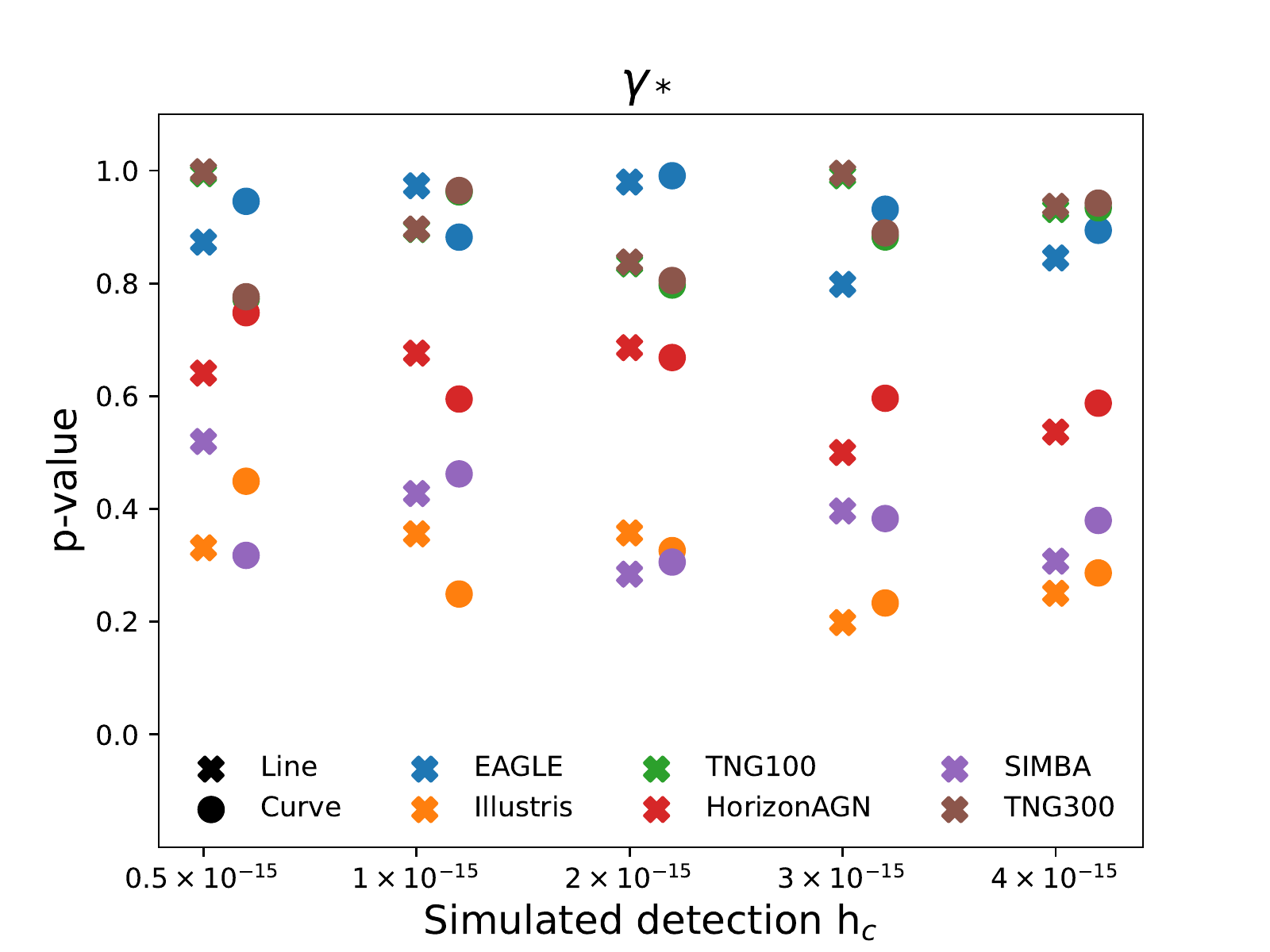}
    \includegraphics[width=0.47\textwidth]{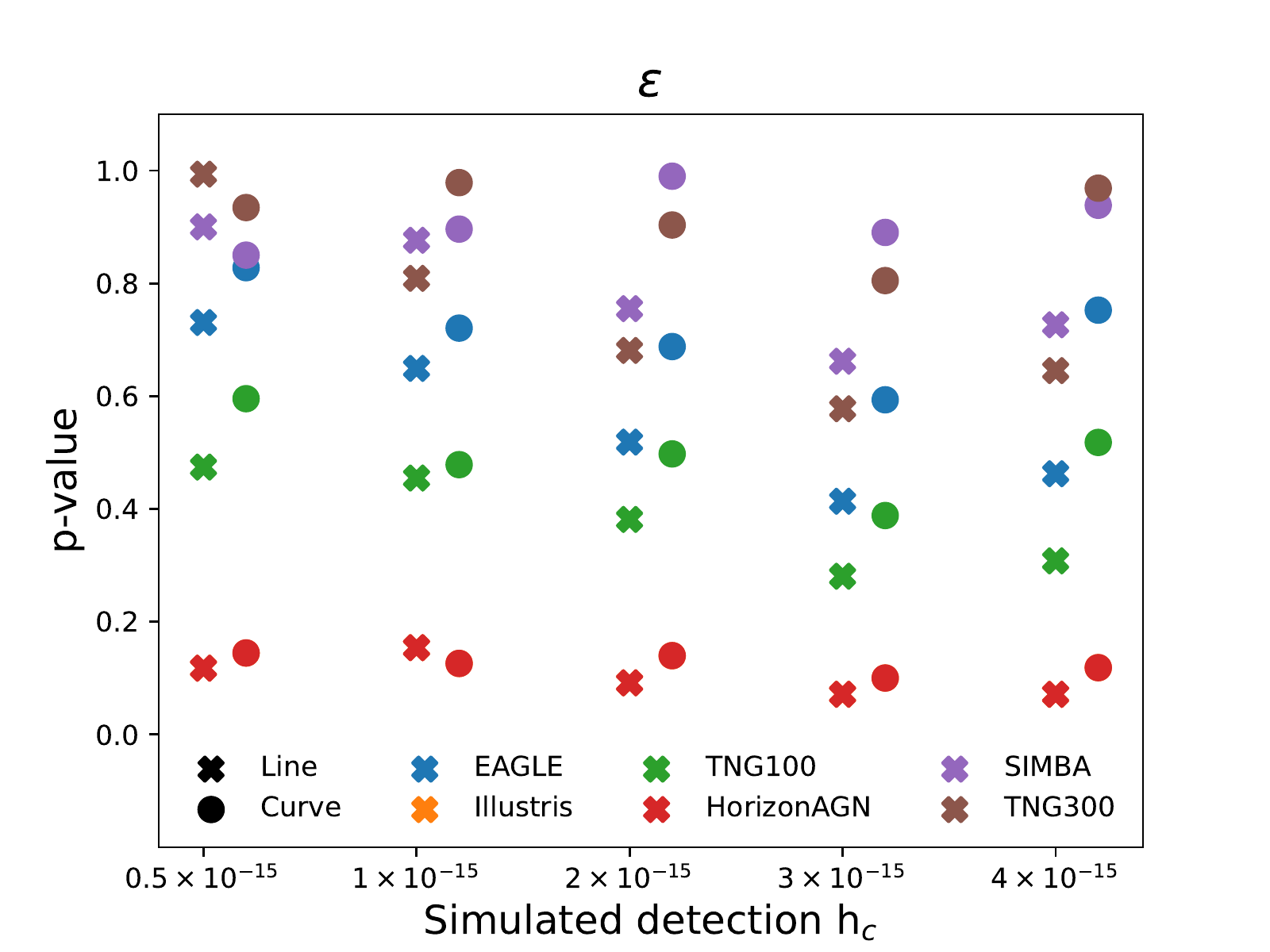}
	\caption{P-values from Kolmogorov-Smirnov tests for the fitted values from the simulations on the 1D marginalized posterior distributions for $(\alpha_*,\beta_*,\gamma_*,\varepsilon)$. Crosses and circles indicate p-values from the straight line and curved spectra simulated detections respectively.} \label{fig:ks_pvals}
\end{figure*}

\begin{figure}
	\centering
	\includegraphics[width=\linewidth]{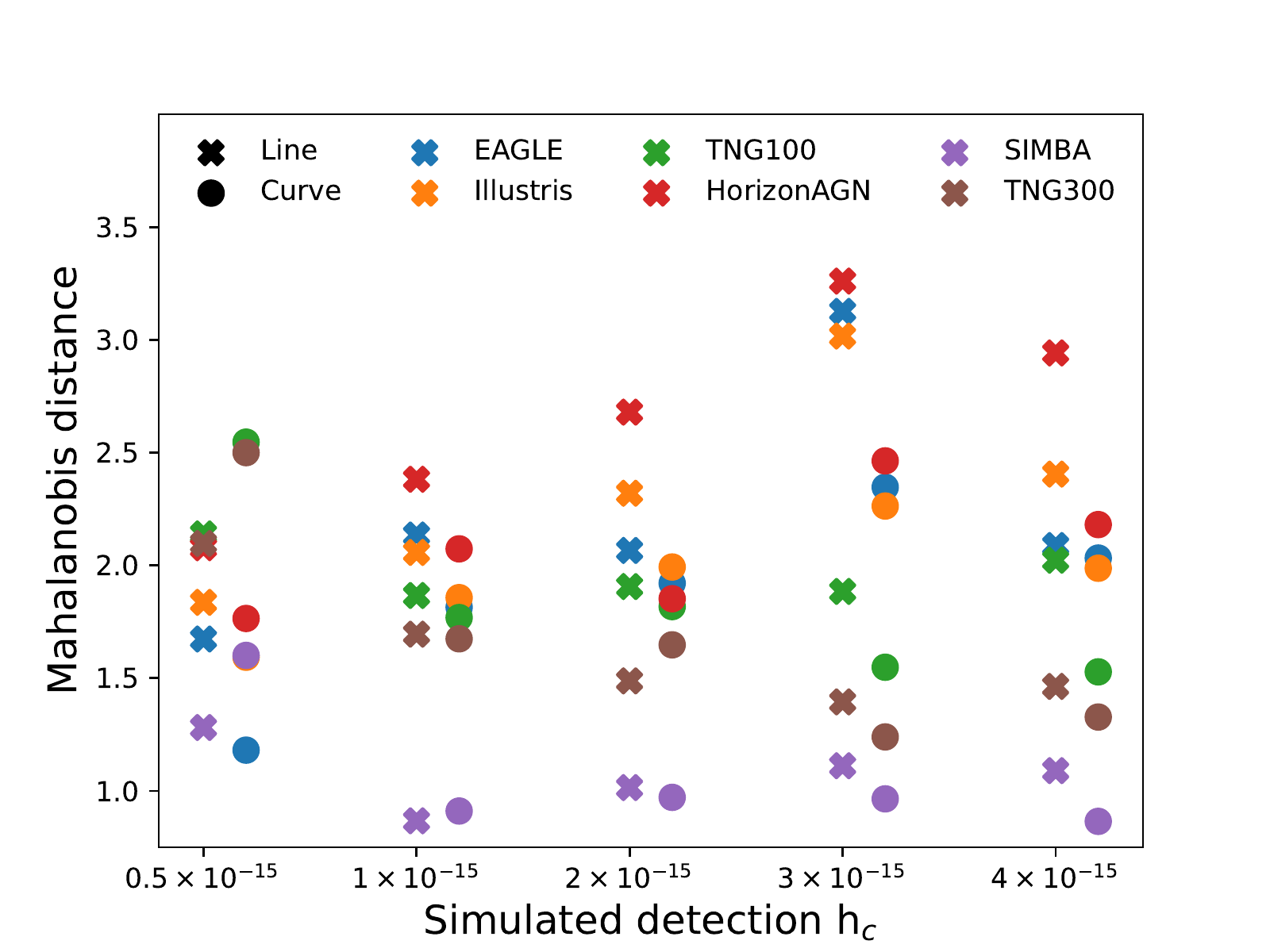}
	\caption{Mahalanobis distances between the fitted values from the simulations and the median values from the posterior constraints for all simulated data sets and both curved (circles) and straight line (crosses) spectra.} \label{fig:mahalanobis}
\end{figure}

\begin{sidewaysfigure*}
	\vspace{-17cm}
	\centering
	\includegraphics[scale=0.4]{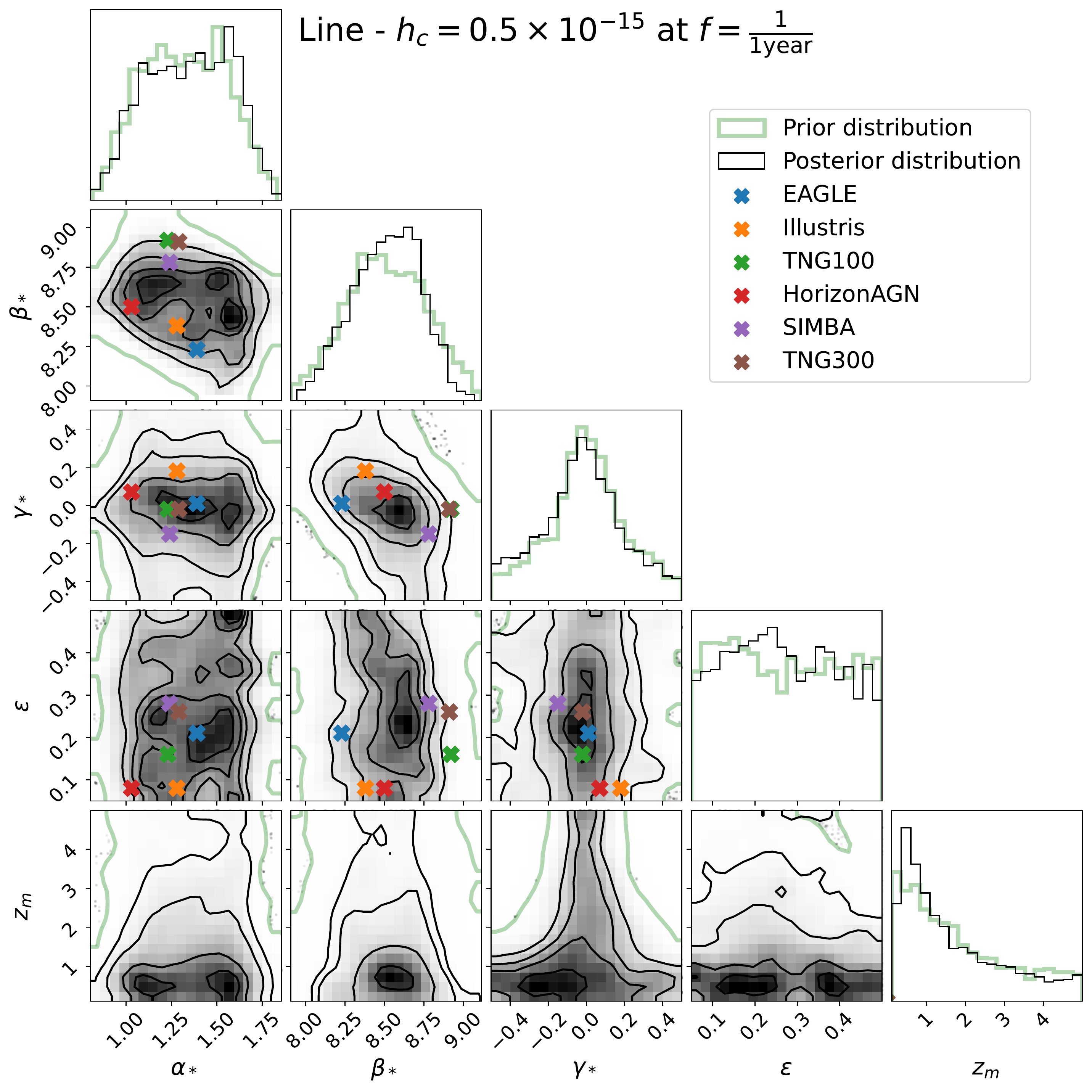}
	\includegraphics[scale=0.4]{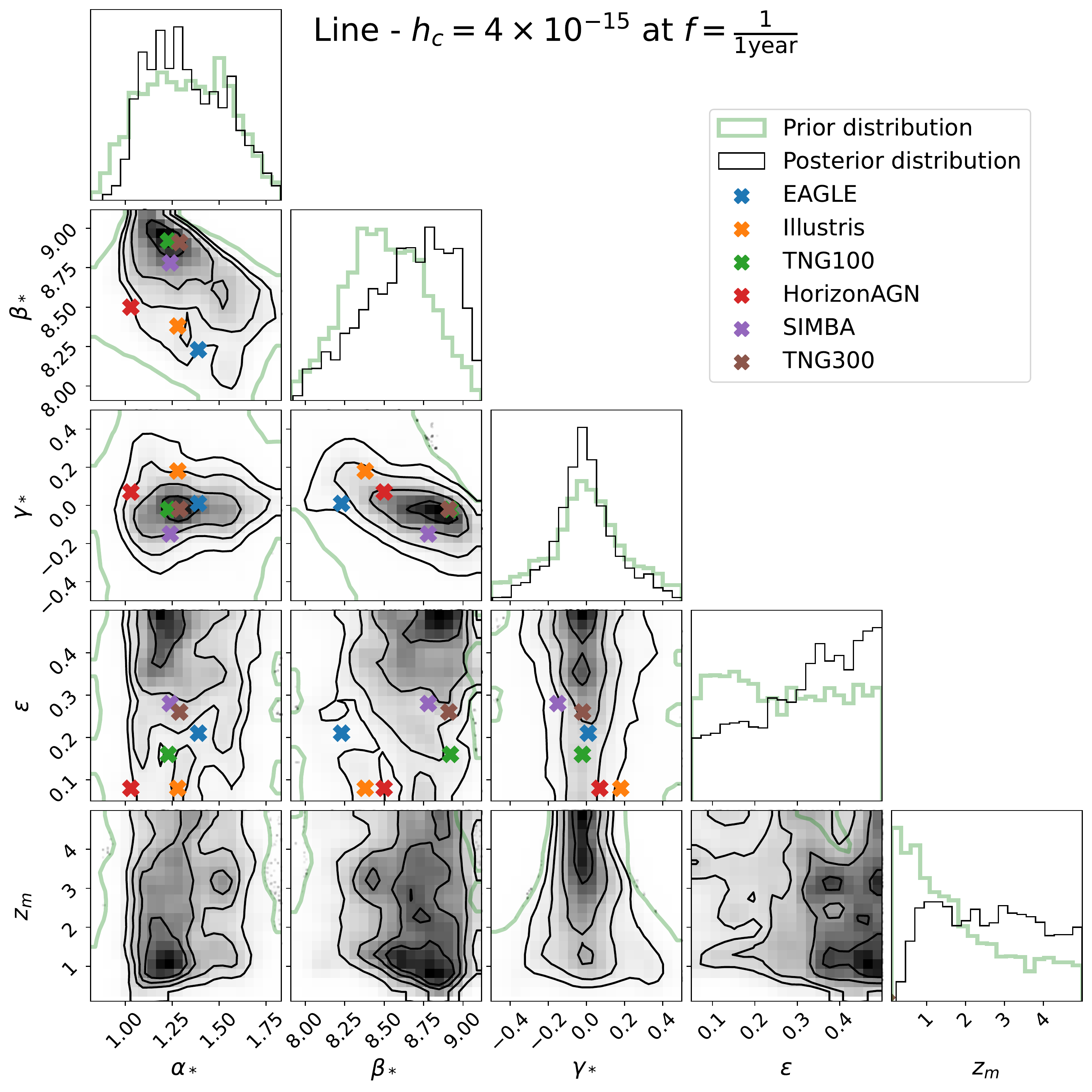}
	\caption{Posterior distributions of $ \alpha_*, \beta_*, \gamma_* $, $\varepsilon$ and maximum redshift $z_m$ for two different straight line characteristic strain spectra of $ 0.5 $ and $ 4 \times 10^{-15} $ at $f=1/1$year are shown as black contours on the left and right panel respectively. The prior distributions are denoted by light green lines. The values of the parameters for the six large-scale cosmological simulation are also shown as coloured crosses, which can be used to judge the consistency between a simulation and the posteriors obtained from a simulated PTA GWB detection.}
	\label{fig:l_0.5-4}
\end{sidewaysfigure*}

\begin{sidewaysfigure*}
	\centering
	\includegraphics[scale=0.4]{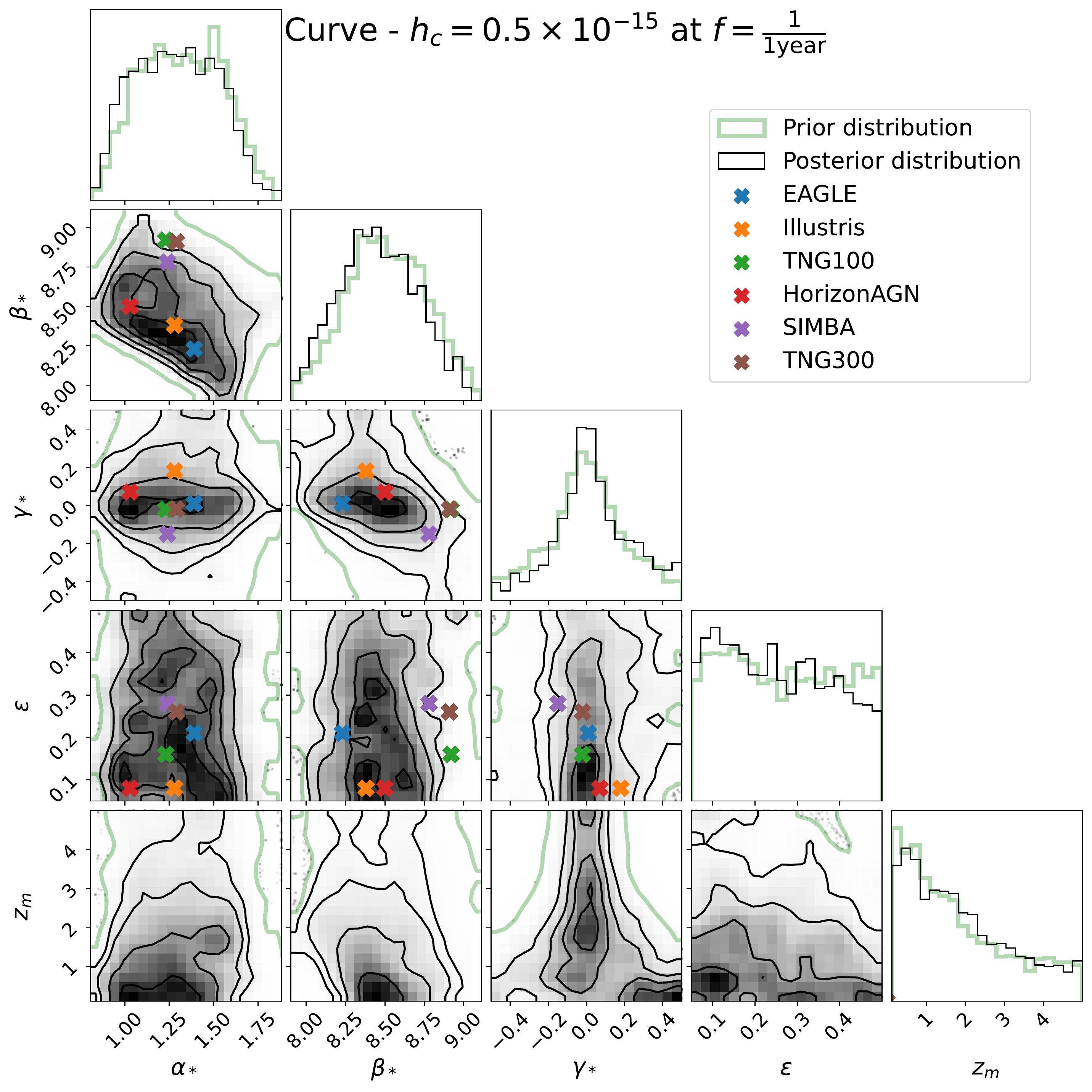}
	\includegraphics[scale=0.4]{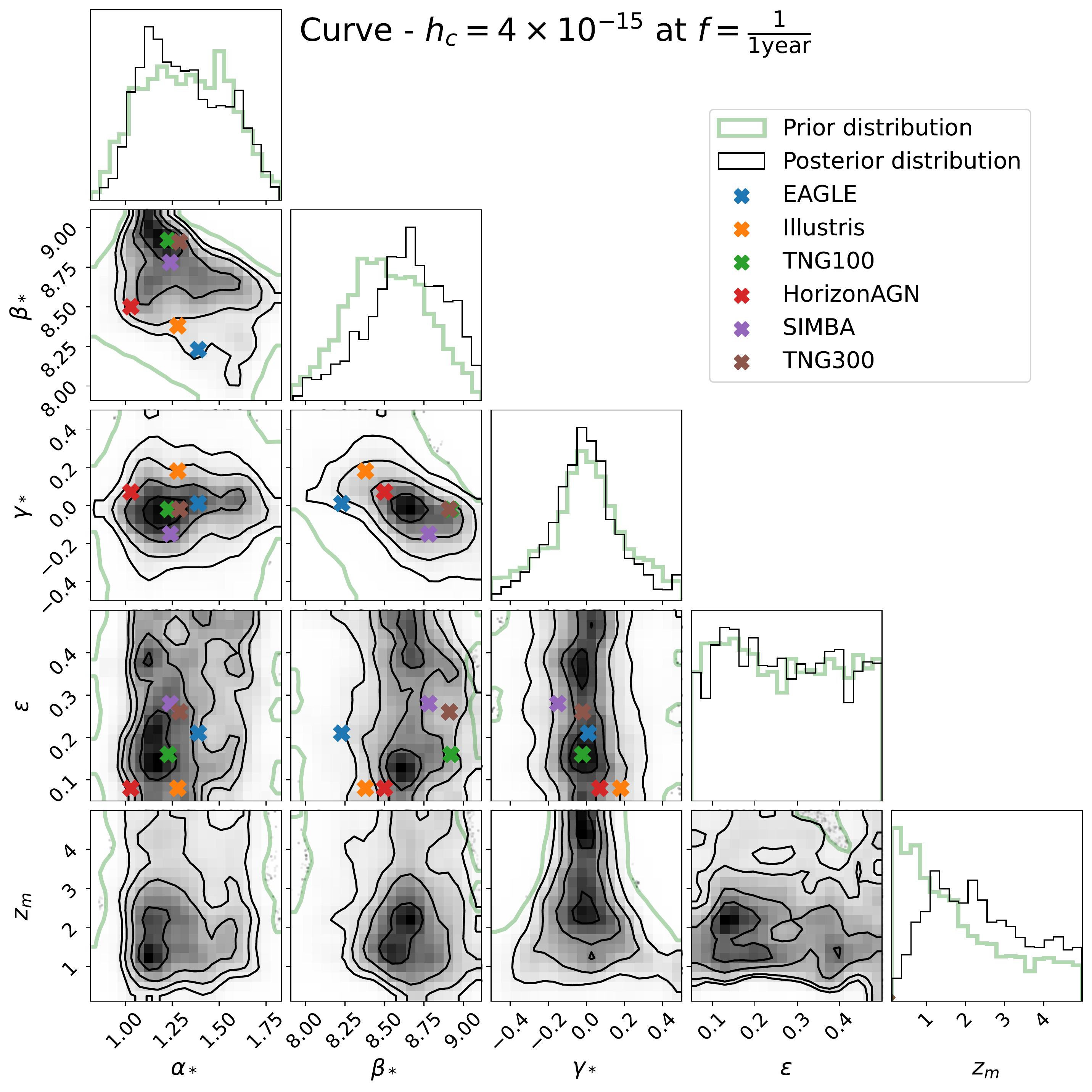}
	\caption{Same style as \autoref{fig:l_0.5-4}, but for two different curved characteristic strain spectra of $ 0.5 $ and $ 4 \times 10^{-15} $ at $f=1/1$year.}
	\label{fig:c_0.5-4}
 \vspace{-17cm}
\end{sidewaysfigure*}

\begin{figure*}
	\centering
	\includegraphics[width=\textwidth]{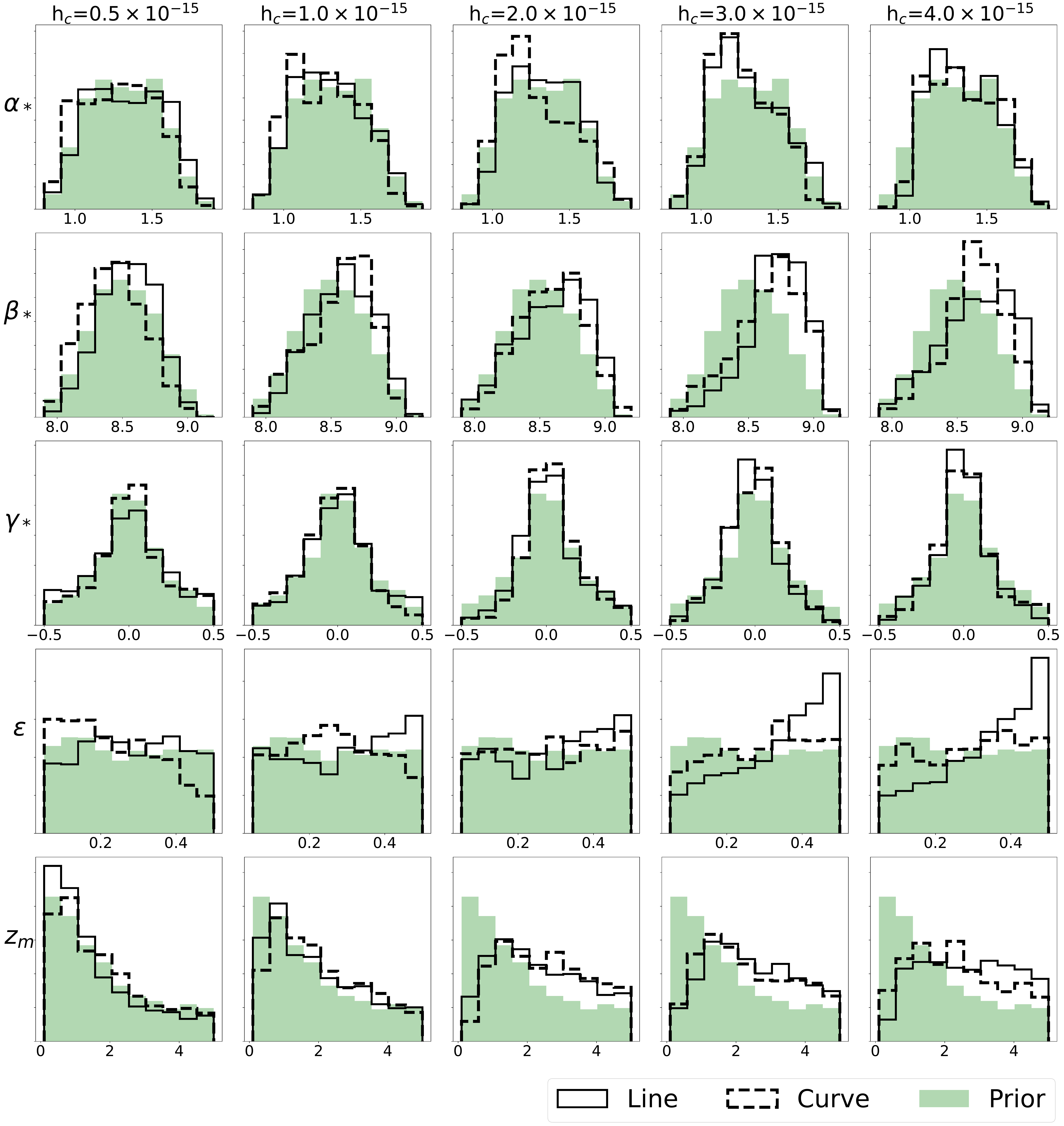}
	\caption{Posterior distributions (in black) of all the BH-bulge mass parameters including redshift variation (in the rows) as the detected characteristic strain value increases (in the columns) for both straight line (solid) and curved (dashed) spectra.For comparison the prior distributions are shown by the green shaded areas.} \label{fig:mbh_step}
\end{figure*}

Simulated PTA GWB detections are used to perform the Bayesian analysis to find the posterior constraints on the astrophysical parameters in our model. We first investigate
the consistency of the fitted values that are an approximate representation of the complex simulations with the different shapes and strains of the simulated PTA detections.

\autoref{fig:ks_pvals} shows the p-values from Kolmogorov-Smirnov (KS) tests on whether the parameters from the six simulations can be consistent with being drawn from the underlying posterior distributions. Since we have four parameters in the redshift dependent BH-bulge relation, each parameter is investigated independently. In general, for most of the simulations and simulated detections, the p-value are well above 0.1, indicating that the fitted values are possible draws from the posterior distributions.
For $\alpha_*$, $\gamma_*$ and $\varepsilon$ the p-values do not vary much for each simulations across the different strains and shapes of the GWB spectrum. The main changes can be seen for $\beta_*$, this could be due to dominant role $\beta_*$ plays in determining the overall GWB strain level. We can very broadly see two trends in the p-values: 1. where they tend to grow as the GWB strain increases and 2. where they behave in the opposite way. Looking at \autoref{fig:ks_pvals} the simulations can be separated by the two trends into two groups: 1. TNG100, TNG300 and SIMBA, following the first trend and 2. EAGLE, Illustris and HorizonAGN, which behave by the second trend.

As the KS tests are performed on marginalized 1D distributions and do not take covariances into account, we also employ the Mahalanobis distance to give another quantity for the consistency between a simulation and a simulated PTA detection. All simulations give distances between about 1 to 3.5 for all GWB strains and spectral shape, see \autoref{fig:mahalanobis}. The same two groups of simulations can be found to follow the same trend, where the first (TNG100, TNG300 and SIMBA) have decreasing distances and the second (EAGLE, Illustris and HorizonAGN) become less consistent. In general, the first look to be more consistent with simulated PTA detection than the second group.

\subsection{Constraints on astrophysical observables}

\subsubsection{BH-bulge mass relation}

\begin{figure*}
	\centering
	\includegraphics[width=\textwidth]{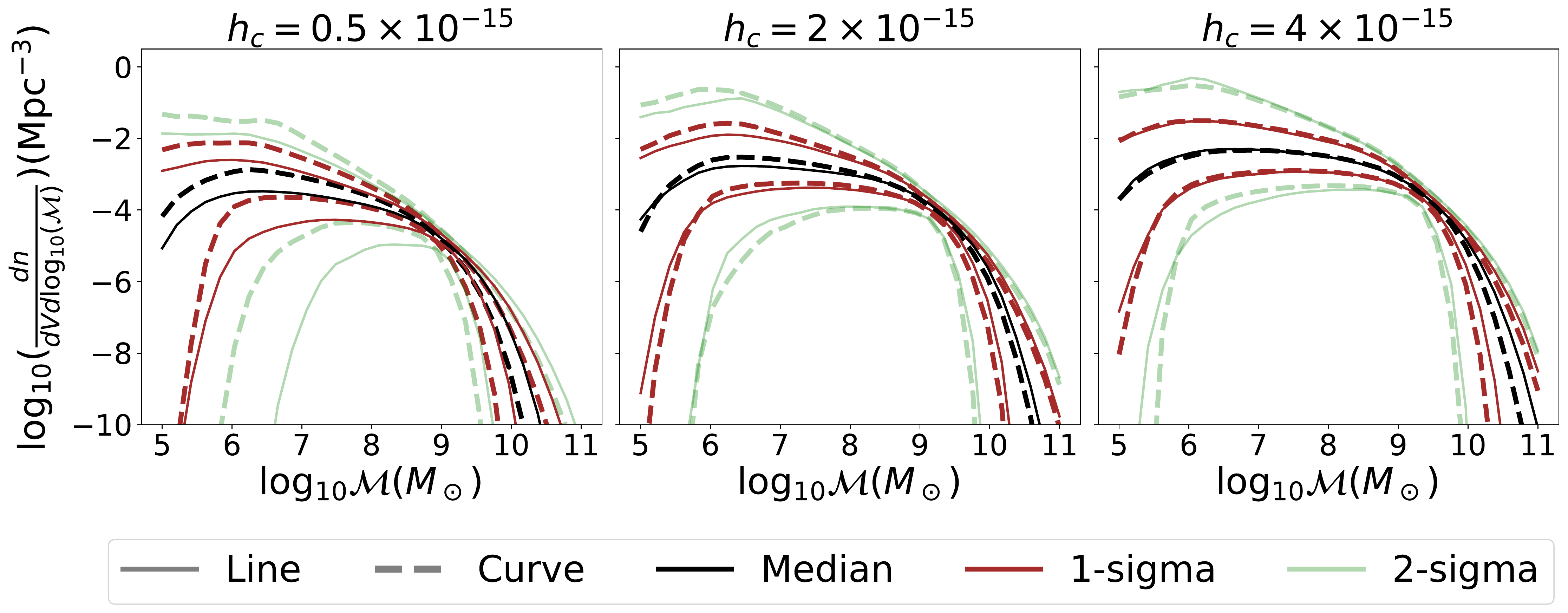}
	\caption{Merger rates with respect to the chirp mass of the SMBHB for increasing characteristic strain values for both straight line (solid) and curved (dashed) spectra computed from the posterior distributions of the Bayesian analysis. The median, central 1 and 2 $\sigma$ ranges are indicated by black, dark red and light green lines respectively.} \label{fig:dndmc}
\end{figure*}

\begin{figure*}
	\centering
	\includegraphics[scale=0.28]{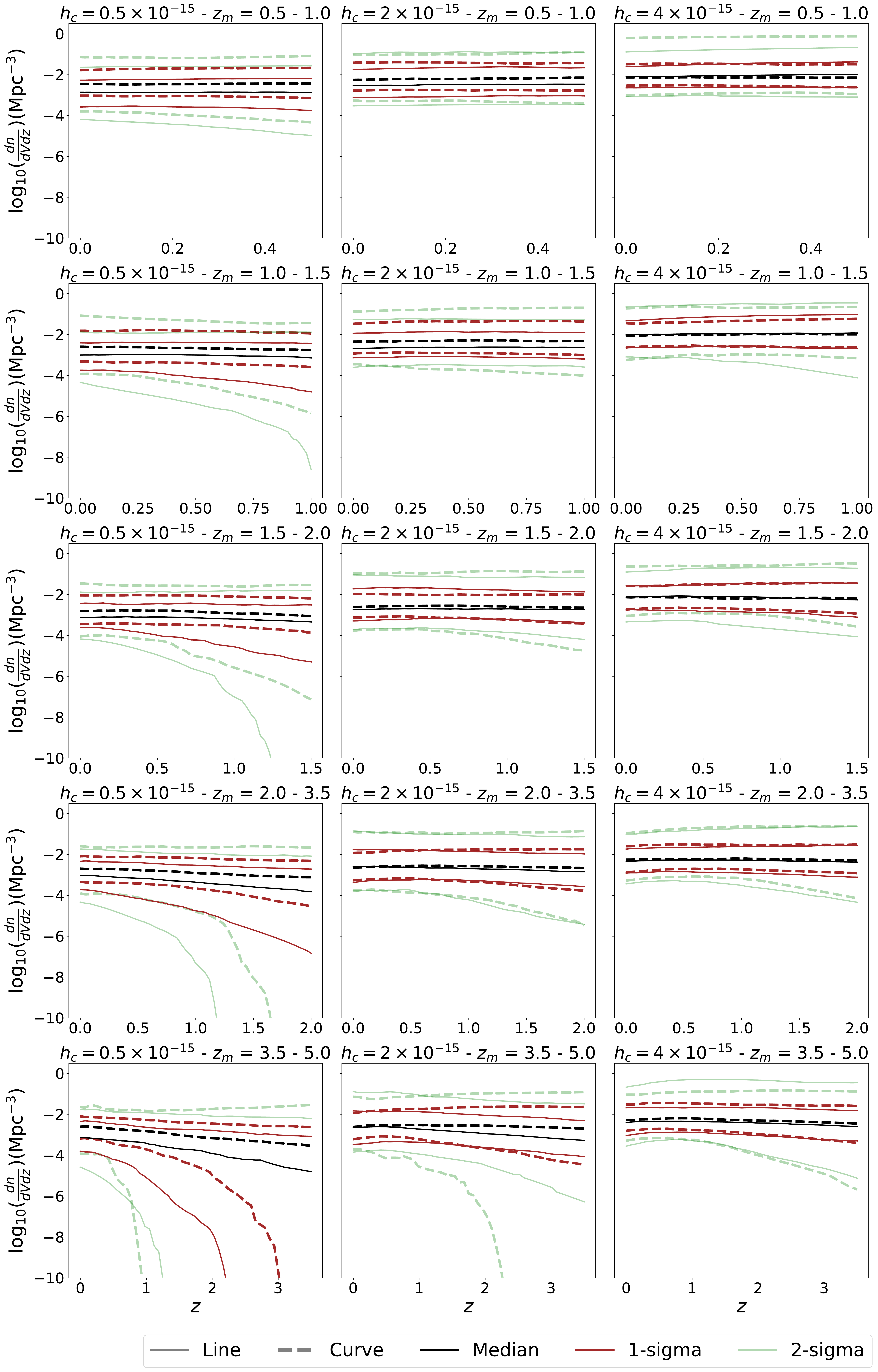}
	\caption{Merger rates with respect to redshift for increasing characteristic strains values for both straight line (solid) and curved (dashed) spectra computed from the posterior distributions of the Bayesian analysis. The median, central 1 and 2 $\sigma$ ranges are indicated by black, dark red and light green lines respectively. Each column represents a GWB detection at a strain amplitude. While each row represents the selection of SMBHBs within a given maximum redshift $z_m$.} \label{fig:dndz}
\end{figure*}

Looking more closely at the posterior constraints on the parameters of the model we note that most of them are very similar to their priors, indicating that they are either already well constrained by other observations or they play only a mild role in the amplitude of the strain values. One of the two main constrained observables is the merger time. It depends on the strength of the GWB, where a higher amplitude leads to shorter merger times and a lower amplitude allows for longer merger times.

The other constrained observable is the black hole - galaxy bulge mass redshift dependent relation, which is why we focus on the parameters $ \alpha_*, \beta_*, \varepsilon$ and $z_m$ in the following. \autoref{fig:l_0.5-4} and \autoref{fig:c_0.5-4} show their 2D and 1D posterior distributions for the cases of circular and eccentric populations respectively. We show only the cases for the smallest and largest amplitudes from our simulated detections.

First, looking at circular population in \autoref{fig:l_0.5-4}, there is little difference between the posteriors (black) and the priors (green) (described in Section \ref{sec:prior}). All simulation fitted values lie within the allowed region. A detected amplitude of $h_c = 0.5 \times 10^{-15}$ provides little extra information. As the amplitude increases to $h_c = 4 \times 10^{-15}$, certain regions of the parameter space are ruled out. Noticeably $\beta_*$ and $\varepsilon$ both show a tendency for larger values. As high redshift black holes tend to be heavier, a trend for faraway SMBHBs also starts to emerge.

Introducing a bend at the lowest frequencies from eccentric population of SMBHBs, shown in \autoref{fig:c_0.5-4}, only marginally changes the findings from circular populations. The eccentricity and the environment of the black holes can only have an effect, if the bend is more prominent in the PTA frequency band.

The evolution of the parameter constraints with amplitude can be found in \autoref{fig:mbh_step}. As the characteristic strain amplitude becomes higher most parameters $ \alpha_*, \beta_*, \varepsilon$ and $z_m$ prefer higher values and $ \gamma_* $ becomes closer to zero. This suggests that a PTA detection can put constraints on the redshift evolution of the BH-bulge mass relation.

Given the PTA detections of a common signal of amplitude $\sim 2.5 \times 10^{-15}$ and the recent evidence for the GW origin, the constraints on our model will be between the two closest matching simulated detections at $2$ and $3 \times 10^{-15}$. However, we use a 25 year observation time span, compared to the $\sim 15$ years of the most recent PTA data sets. Our model also takes into account for the possibility that the BH-bulge relation could evolve with redshift and samples for the maximum redshift, which is equivalent to the volume of space, that PTAs can constrain. Extensive astrophysical interpretation was performed by EPTA \citep{epta_interp} and NANOGrav \citep{ng15_astro}, which are consistent with our findings.

The corner plots for the complete 20 parameters with amplitudes  $h_c = 0.5 \times 10^{-15}, 1\times 10^{-15}, 2\times 10^{-15}, 3\times 10^{-15},$ and $4\times 10^{-15}$ for both circular and eccentric population of SMBHBs are presented in Appendix C.

\begin{figure*}
	\centering
	\includegraphics[width=\textwidth]{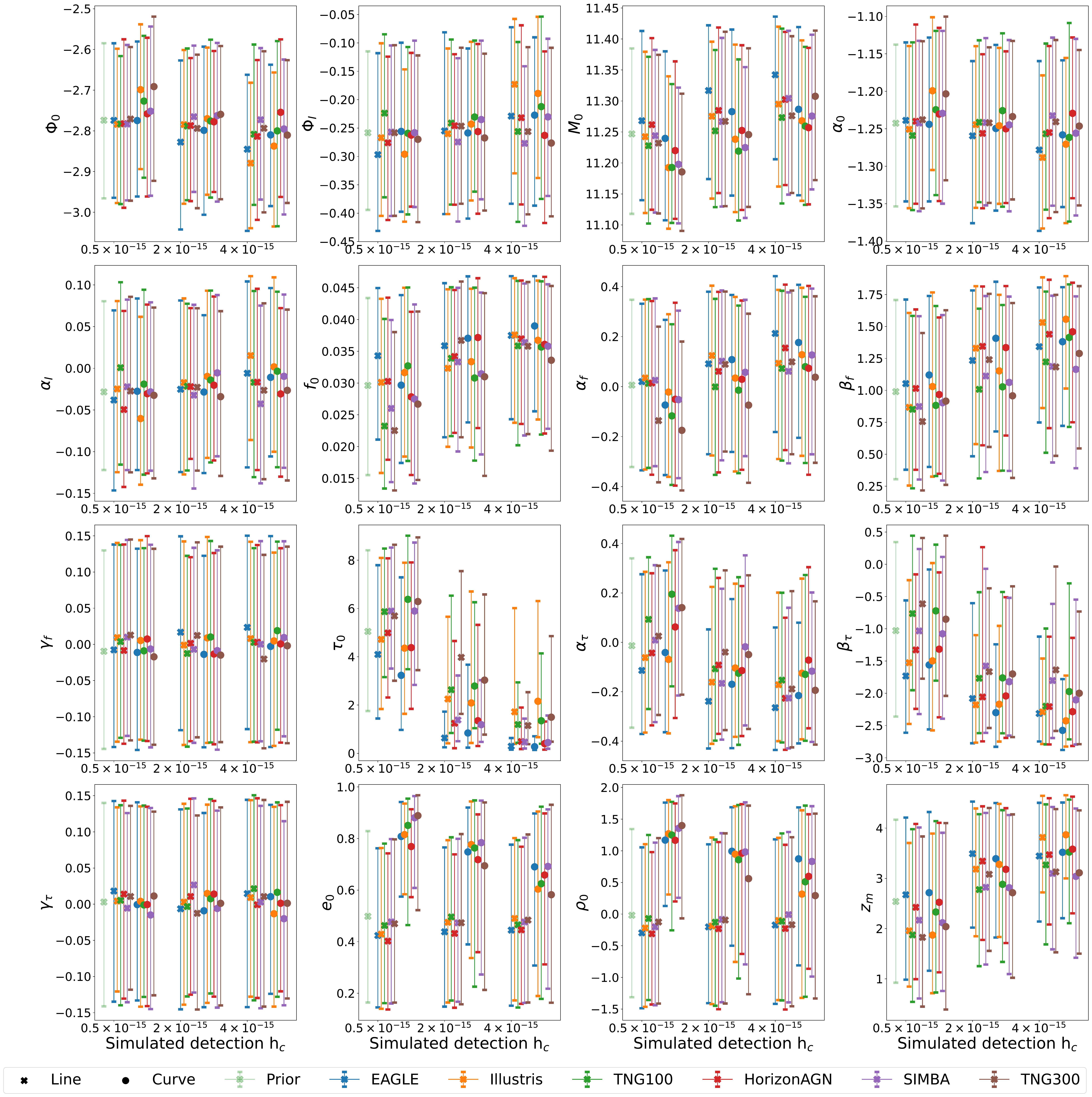}
	\caption{Median values and central 68\% credible regions of the 1D marginalized posteriors from the 16 parameters using the best fit values for the BH-bulge relation for the six simulations and three different strain values and shapes. The prior distributions are indicated by the leftmost light green point in each panel. Crosses indicate straight line spectra, while circles show curved spectra. A set of six coloured points represent the results from one simulated detection case for all six simulations.} \label{fig:unc}
\end{figure*}

\begin{figure*}
	\centering
	\includegraphics[scale=0.111]{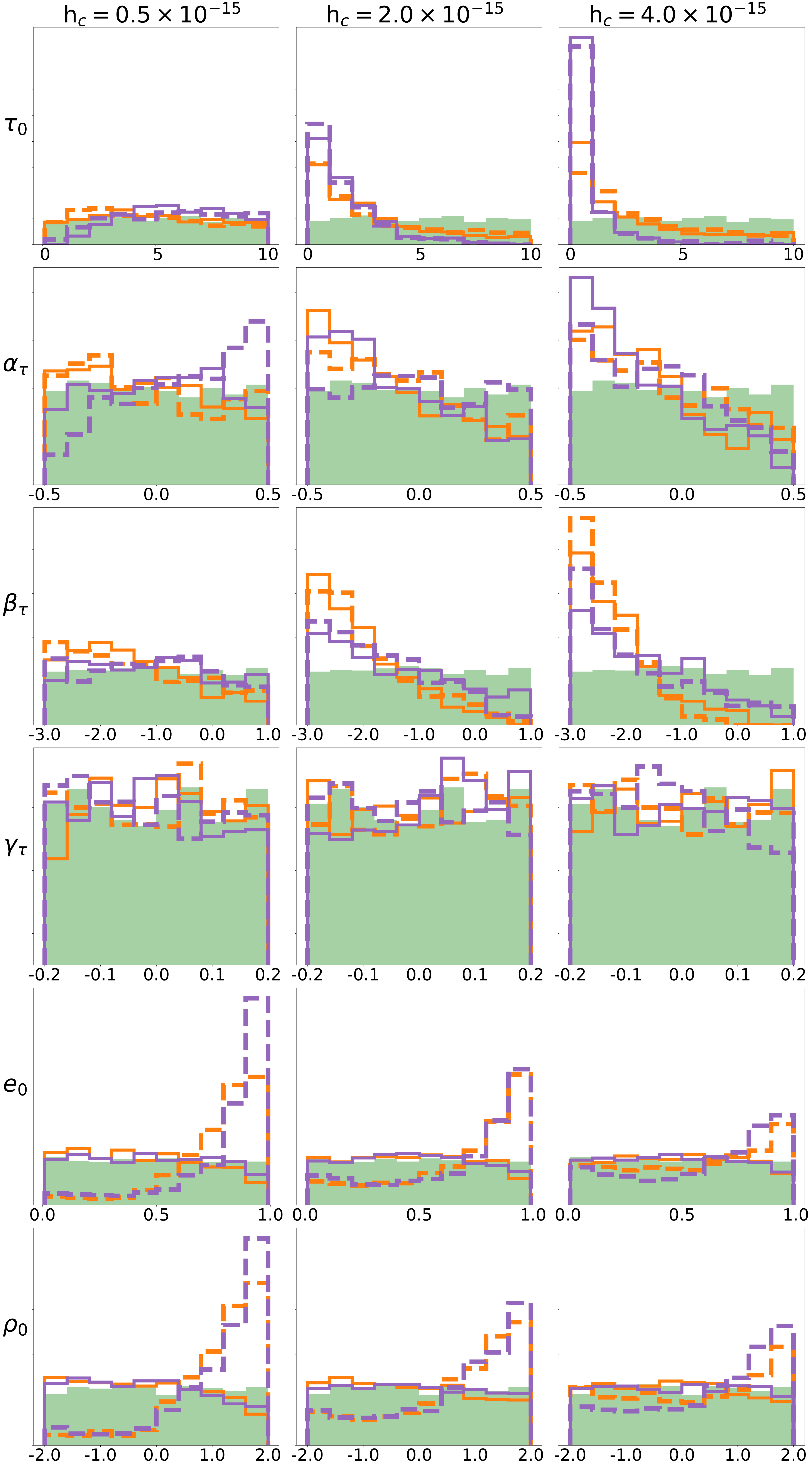} \ \quad
	\includegraphics[scale=0.111]{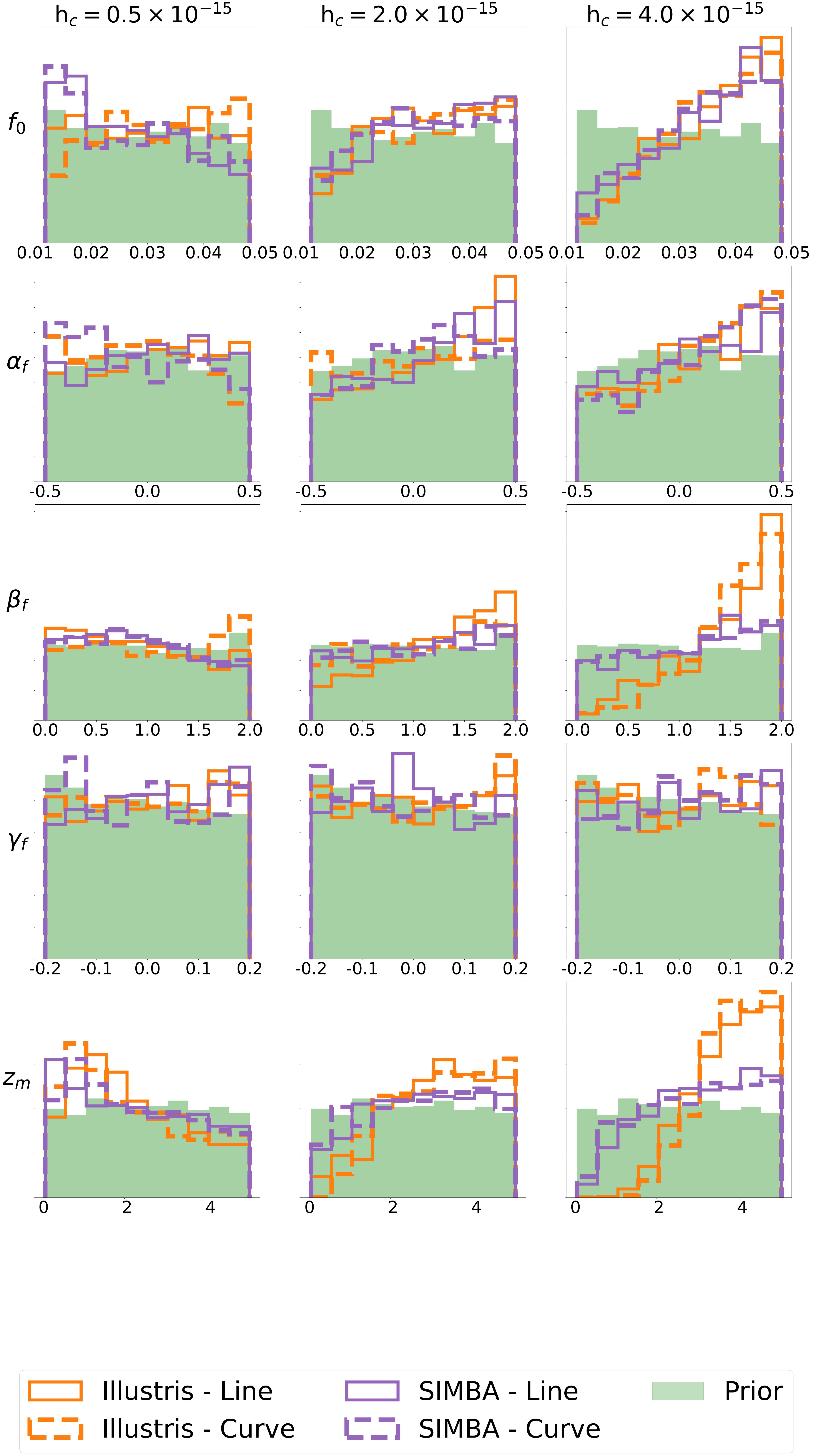}
	\caption{Posterior distributions of selected astrophysical parameters for both straight line (solid) and curved (dashed) characteristic spectra of increasing strain with BH-bulge mass parameter values fixed to those representative of Illustris and SIMBA. The green shaded areas indicate the prior distributions for comparison.}
 \label{fig:11para}
\end{figure*}

\subsubsection{SMBHB merger rate}

An interesting quantity that can be computed from our model is the merger rate of the SMBHBs from Eqn \eqref{eqn:BH_mergerrate} given the constraints on the parameters from the simulated GWB detections. Following \cite{2019MNRAS.488..401C}, we first integrate over the mass ratio, leaving a merger rate by redshift and chirp mass. Next, we can integrate over the mass and redshift to get \autoref{fig:dndmc} and \autoref{fig:dndz} showing $\dd n/\dd \mathcal{M}$ and $\dd n / \dd z$ respectively.

The merger rates with respect to the SMBHB mass in  \autoref{fig:dndmc} are very similar between the circular and eccentric populations at most investigated strain amplitudes. This indicates that environmental effects are not strongly covariant with the population properties. Only at the lowest amplitude $h=0.5\times10^{-15}$ differences become noticeable with the eccentric population having a larger number of low mass binaries and the high mass drop-off at lower masses, compared to the circular population. This general trend persists through increasing amplitudes, but becomes less significant. In general, with larger amplitude the rate of massive binary mergers also grows. The median merger rate moves towards a drop-off at higher mass. Additionally, one can see an increase of the merger rate for smaller mass binaries in the 2-sigma range, especially in the circular population.

As we introduced the maximum redshift as a free parameter, the merger rates with respect to the redshift drop to zero at different maximum redshifts. This mimics the expectation that the GWB that PTAs are sensitive to will be dominated by closeby binaries. As such, we have binned the posterior samples by their maxmimum redshift. Within each bin we plot the merger rate within a common range of redshifts in \autoref{fig:dndz}. E.g., in the left most column we selected all the posterior samples with a maximum redshift of $z<1$ and plot the integrated merger rate between 0.1 and 0.5.

In general, the median merger rate as a function of redshift is nearly constant across most redshifts, amplitudes and different populations. A small raise as the detected amplitude increases can be seen. The difference between the two populations is very small with the eccentric population requiring an overall larger number of mergers. As in the mass dependent merger rates, the main differences can be seen at the lowest amplitude. At $h=0.5\times10^{-15}$ the drop of the lower bounds of the merger rate at high redshift is clearly visible. This is consistent with the prior assumption of a possible decreasing number of SMBHBs at high redshifts contributing to the GWB. Consequently, the number of samples for the highest maximum redshift is also low. If a detection favours high amplitudes, more binaries even at large distances are required to produce the GWB. This can be seen most prominently in the rightmost column in  \autoref{fig:dndz}, where the 2-sigma lower bound drop of the merger rate moves from $z\approx1$ to $z\approx4$.

\subsection{Constraints from simulations}

By fixing the BH-bulge mass parameters to the best fit values from the simulations given in \autoref{tab:sims}, we can see how the constrains on the other parameters are affected. For computational cost reasons we only analyze the $h_c = 0.5\,, 2\,, 4 \times 10^{-15}$ detections for both straight line and curved GWB spectra. \autoref{fig:unc} shows the median values and central 68\% of the 1D marginalized posterior distributions for all six simulations.

In general, most parameters have similar posterior compared to the prior constraints (in light green). The five parameters related to the GSMF $(\Phi_0\,, \Phi_I\,, M_0\,, \alpha_0\,, \alpha_I)$ are already well constrained from observations. Parameters that play only a subdominant role, like those for the pair fraction $(f_0\,, \alpha_f\,, \beta_f\,, \gamma_f)$ and maximum redshift $z_m$ are only slightly constrained towards larger values for stronger GWB strains. The eccentricity $e_0$ and stellar density $\rho_0$ parameters are degenerate. However, we can see that straight line spectra result in low eccentric binaries in low stellar dense environments, while a curved spectrum indicates the need for either eccentricity or dense stellar environments of the binaries. Lastly, the most important observable when using a fixed BH-bulge relation is the merger time. A short merger time is needed to produce a stronger GWB, especially if the masses of the SMBHs are fixed to results from cosmological simulations. The (second column, third row) panel in \autoref{fig:unc} on the merger time norm $\tau_0$ shows for all six simulations this decrease of the median values as well as the shrinkage of central 68\% credible regions. The other three parameters describing the merger time $(\alpha_\tau\,, \beta_\tau\,, \gamma_\tau)$ play a minor role and are thus not much more constrained compared to the prior.

The corner plots for the 16 parameters with amplitudes  $h_c = 0.5 \times 10^{-15}, 2\times 10^{-15},$ and $4\times 10^{-15}$ for both circular and eccentric population of SMBHBs using the fitted BH-bulge mass parameters from the simulations can be found in the online supplementary material.

\begin{figure*}
	\centering
	\includegraphics[width=\textwidth]{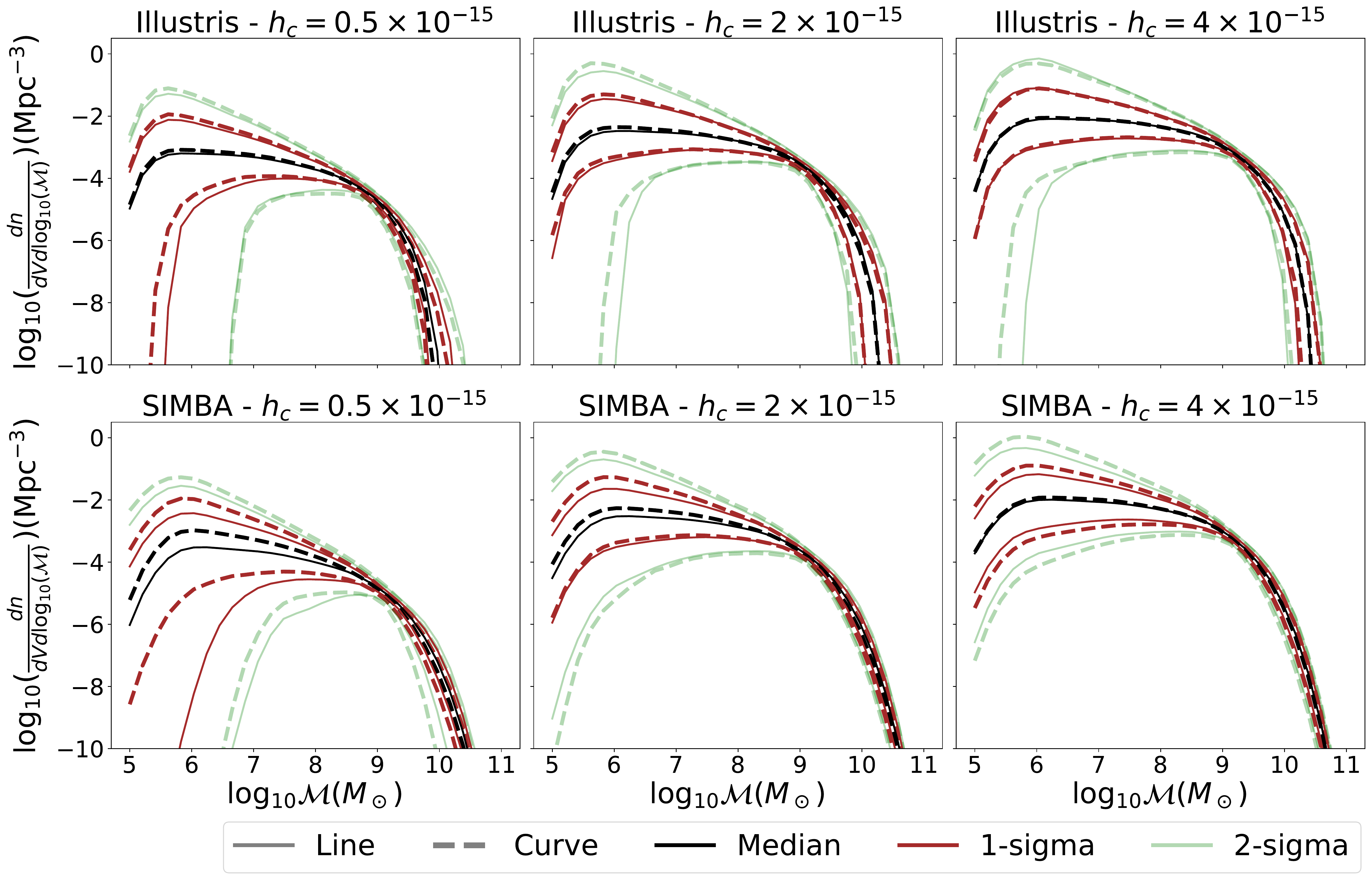}
	\caption{Same style as \autoref{fig:dndmc}, but for merger rates with respect to the chirp mass of the SMBHB using values fixed to Illustris and SIMBA in the Bayesian analysis.} \label{fig:sim_dndmc}
\end{figure*}

\begin{sidewaysfigure*}
    \vspace{-17.7cm}
	\centering
	\includegraphics[scale=0.22]{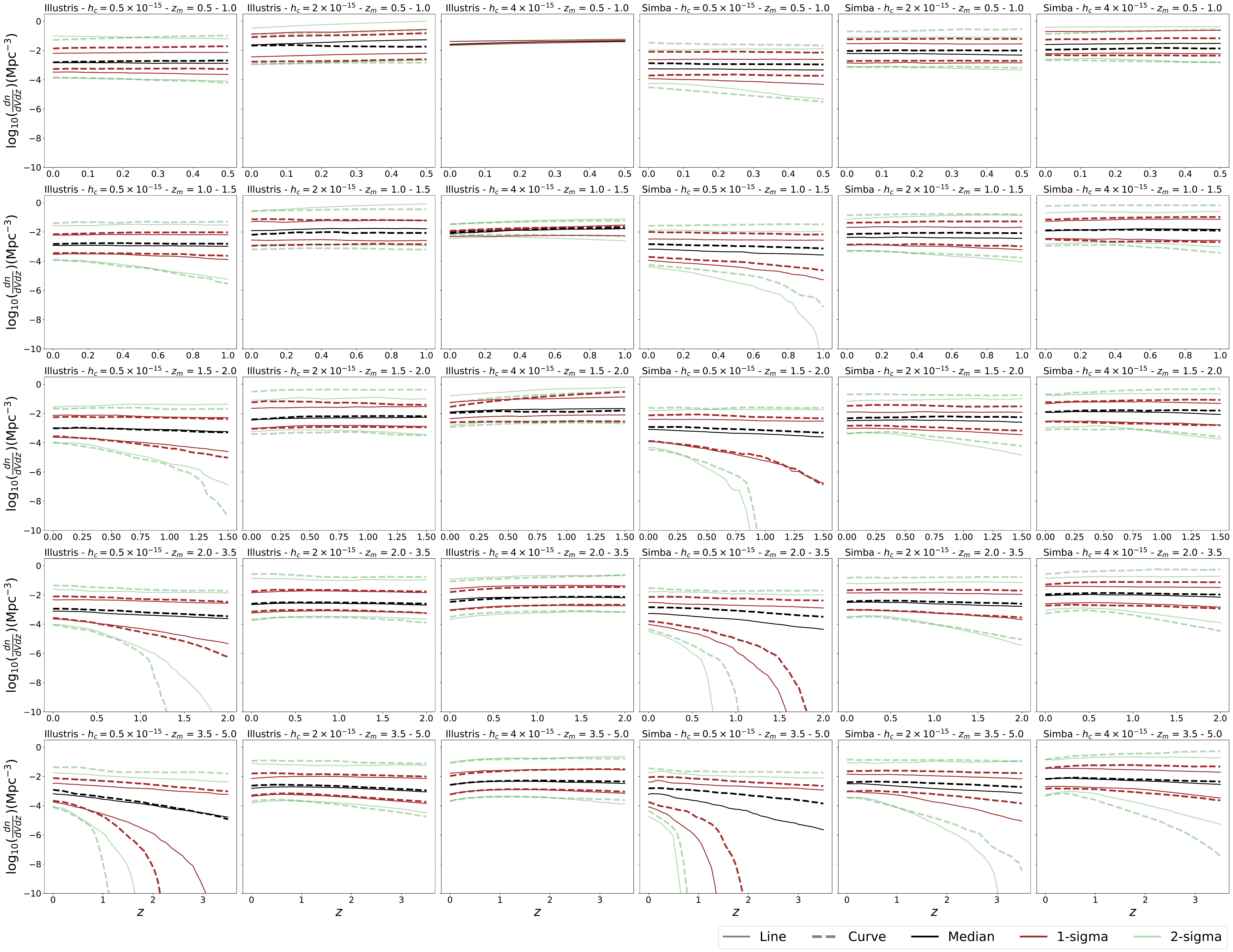}	
	\caption{Same style as \autoref{fig:dndz}, but for merger rates with respect to redshift using values fixed to Illustris and SIMBA in the Bayesian analysis. The left three columns are for Illustris, whereas the right three columns are for SIMBA.} \label{fig:sim_dndz}
\end{sidewaysfigure*}

\subsubsection{Parameter constraints from Illustris and SIMBA}

Below, we focus on constraints from the Illustris and SIMBA simulations.
These two cosmological simulations are chosen since Illustris shows positive while SIMBA shows negative evolution of the SMBH mass with redshift, and thus are the extreme two cases in these six large-scale cosmological simulations. The distribution of all the astrophysical parameters for both curved and straight line characteristic spectra of the GWB and with fixed BH-bulge mass parameter values to match Illustris and SIMBA are given in the \autoref{fig:11para}, where Illustris is shown in orange, SIMBA in purple and the prior of the parameters in the green shaded regions.

The evolution of the pair fraction parameters displays similarities in both the curved and straight line characteristic spectra for different amplitudes in Illustris and SIMBA, with two noticeable differences: 1. $f_0$ at $h_c = 0.5 \times 10^{-15}$ and 2. $\beta_f$ at $h_c = 4 \times 10^{-15}$. At a characteristic spectrum value of $h_c = 0.5 \times 10^{-15}$, both Illustris and SIMBA exhibit posteriors similar to the prior for the pair fraction norm $f_0$. However, at this strain amplitude, while Illustris trends towards larger values, SIMBA behaves in the opposite way. As the strain value increases, both Illustris and SIMBA start to display posterior distributions that prefer larger values of $f_0$.
The pair fraction mass slope $\alpha_f$ for both Illustris and SIMBA shows a preference for low values at $h_c = 0.5 \times 10^{-15}$, followed by no preference at $h_c = 2 \times 10^{-15}$, and then higher values at $h_c = 4 \times 10^{-15}$ for both the curved and straight line spectra. The posterior distributions of the pair fraction redshift slope $\beta_f$ exhibit similar evolution with amplitude as in the case of $\alpha_f$ for both simulations, except at the largest strain value, where the trend is more pronounced in Illustris compared to SIMBA.
In conclusion, the pair fraction increases with larger amplitudes for both circular and eccentric populations. More massive and distant galaxy pairs are required to produce the gravitational wave background at higher strains. Illustris tends to require more pairs than SIMBA for the same amplitude.

The curved characteristic spectra with values of $h_c = 0.5\,, 2 \times 10^{-15}$ in SIMBA reveal a correlation between higher posterior values of $\alpha_\tau$ and higher values of eccentricity $e_0$ and $\rho_0$.
This is in contrast to the general behaviour that increasing characteristic spectrum values lead to lower values of $\alpha_\tau$, $\beta_\tau$ and $\tau_0$. The posteriors change from being in broad agreement with the priors at $h_c = 0.5 \times 10^{-15}$ for both curved and straight line spectra and both simulations to trending very clearly towards lower merger times at $h_c = 4 \times 10^{-15}$. The main difference between Illustris and SIMBA seems to be that the merger time redshift slope $\beta_\tau$ is more constrained for Illustris, whereas it is the merger time norm $\tau_0$ for SIMBA. We can see that the curved and straight line spectra at the same amplitude mostly impact the eccentricity $e_0$ and stellar density $\rho_0$ parameters. The straight line spectra lead to almost no constraints at all strains. On the other hand, the curved line spectra show the correlation between these two parameters in creating a bend at low GW frequencies. It should be noted that the posteriors look to be less well constrained for $e_0$ and $\rho_0$ with larger amplitudes in the curved spectra case. This could be from the difficulties of PTA detections to accurately measure a bend in the GWB spectrum, especially for our simulated detections with limited frequency coverage.

Finally, the inclusion of the maximum redshift $z_m$ parameter allows to gauge where the most dominant SMBHBs can be found for a simulated PTA GWB detection and a chosen cosmological simulation. The last row in \autoref{fig:11para} shows that in general larger strain values require binaries to be concentrated at higher redshifts. For both the curved and straight line spectra Illustris constraints more strongly to large maximum redshifts, while SIMBA only shows a weak trend in the same direction. This could be the effect of the $\gamma_*$ parameter that describes the BH-bulge relation, where positive values, like in Illustris, produce more massive BHs at higher redshifts. Whereas the negative value in SIMBA leads to the most massive BHs being at smaller redshifts.

\subsubsection{Mergerrate constraints of Illustris and SIMBA}

It is interesting to look at $\dd n/\dd \mathcal{M}$ and $\dd n/ \dd z$ as in the previous section for the Illustris and SIMBA simulations as shown in  \autoref{fig:sim_dndmc} and \autoref{fig:sim_dndz}.
The most prominent feature of $\dd n/\dd \mathcal{M}$ in \autoref{fig:sim_dndmc} is a shift towards larger mass from Illustris to SIMBA for the same GWB strain. This is consistent with the prediction that SIMBA produces lower mass binaries than Illustris and thus need more binaries to match the emitted GWB strain.
The free parameters are adjusted to get same amplitude as explained above from  \autoref{fig:11para}. The other feature that is visible from  \autoref{fig:sim_dndmc} is the variation between circular and eccentric binaries producing a straight and curved GWB strain spectrum respectively. There is no difference in the median of the merger rates with respect to the SMBHB mass in Illustris for the circular and eccentric binaries with same amplitude, however we can see small differences at the lower 2-sigma boundaries. SIMBA clearly shows a slight variation in binary chirp mass between circular and eccentric binaries.

The merger rate with respect to the redshift is shown in  \autoref{fig:sim_dndz} for Illustris and SIMBA with the panels defined in the same way as in the previous section and  \autoref{fig:dndz}. An important feature here is that no GWB strain amplitude of $h_c = 4 \times 10^{-15}$ can be obtained from a circular population of binaries with redshifts $z<1.0$ and only very few eccentric populations could produce such amplitude in our sampling. Within the small number statistical uncertainties it seems that such a large amplitude is rarely achieved by any simulation within $z<1.0$.
While in general the results from Illustris and SIMBA in \autoref{fig:sim_dndz} are very similar to those in \autoref{fig:dndz}, the merger rates for Illustris become nearly constant across all redshifts at amplitude of $h_c =2 \times 10^{-15}$ already.

\section{Discussion and conclusions}
\label{sec:Conclusions}

The parametric astrophysical model presented in this work describes the intensity of the gravitational wave background as a function of the frequency. The focus was on the redshift dependent BH-bulge mass relation.
By understanding the processes and relationships concerning the formation and co-evolution of galaxies and their central black holes, we have used an analytical expression in order to refine current astrophysical models. This allowed us to compare the predictions of this model with the constraints from PTA observations.

Large-scale cosmological simulations help us to study the evolution of the Universe since observational unbiased data is hard to produce.
We have fitted our redshift dependent BH-bulge relation to a suite of six simulations: EAGLE, Illustris, TNG100, TNG300, Horizon-AGN and SIMBA. The obtained best fit parameters serve as representative values for a Bayesian analysis. In general, all six simulations are consistent within $\leq 3.5 \sigma$ with the range of shapes and strains of our simulated PTA GWB detections. The simulations can be broadly separated into two groups: 1. TNG100, TNG300 and SIMBA, which become more consistent with PTA detection as the GWB increases in amplitude and 2. EAGLE, Illustris and HorizonAGN, which behave in the opposite way. This separation coincidentally also follows the sign of the fitted $\gamma_*$ values of these simulations.

We simulated PTA detections to see how much they can help to constrain the posteriors of the parameters of the redshift dependent BH-bulge mass relation. As the redshift increases the value of $\gamma_*$ becomes more restricted.
We find the tightest constraints for $\beta_*$ from a GWB detection in the PTA range, while $\alpha_*$ does not change much from the prior.

Varying the maximum redshift parameter in the model seen in \autoref{fig:gwb_para_variation} shows that the dominant fraction of the SMBHB population can be found withing $z_m \sim 1.5-2.5$ with the SMBHBs at higher redshifts only contributing a small, but not negligible, amount to the GWB. The study of higher redshift galaxies will be useful to determine the redshift evolution of BH-bulge mass relation. There still are difficulties to observe higher redshift galaxies.

Our proposed BH-bulge mass relation is a first-order extension of the standard redshift independent linear scaling relation. It can fit the masses from the simulations while maintaining approximately constant values of $\alpha_*$, $\beta_*$, $\gamma_*$ and $\varepsilon$ for redshifts $z\leq5$. The results depend on the specific parametric function and thus can only approximate the complexity of the masses given by the simulations.

Additionally, \citep{graham12} propose a double power law for the redshift independent BH-bulge mass relation of galaxies using observational data.
Further studies to find and test the optimal shape of a redshift dependent BH-bulge mass relation are required.

Another interesting area for further improvement of the model is the galaxy stellar-bulge mass relation. The phenomenological stellar-bulge mass relation we have used is more suitable for elliptical and spheroidal galaxies, so a relation containing spiral galaxies including a degree of the spirality will be ideal to study a wide range of galaxies and their central black holes.

\section*{Acknowledgements}
We would like to thank Marta Volonteri for helpful discussions. We acknowledge the support of our colleagues in the European Pulsar Timing Array. MH acknowleges support from the Gliese fellowship of the Zentrum für Astronomie and the MPIA fellowship of the Max-Planck-Institut für Astronomie.  AS acknowledges the financial support provided under the European Union’s H2020 ERC Consolidator Grant ``Binary Massive Black Hole Astrophysics'' (B Massive, Grant Agreement: 818691). We acknowledge financial support from “Programme National de Cosmologie and Galaxies” (PNCG), and “Programme National Hautes Energies” (PNHE) funded by CNRS/INSU-IN2P3-INP, CEA and CNES, France. We acknowledge financial support from Agence Nationale de la Recherche (ANR-18-CE31-0015), France.

\section*{Data availability}
The data used in this article shall be shared on reasonable request to the corresponding authors.

\bibliographystyle{mnras}
\bibliography{bibliography}

\appendix
\section{GWB characteristic spectra from variations of the astrophysical parameters}
\label{sec:appA}
\begin{figure*}
	\centering
	\includegraphics[width=\textwidth]{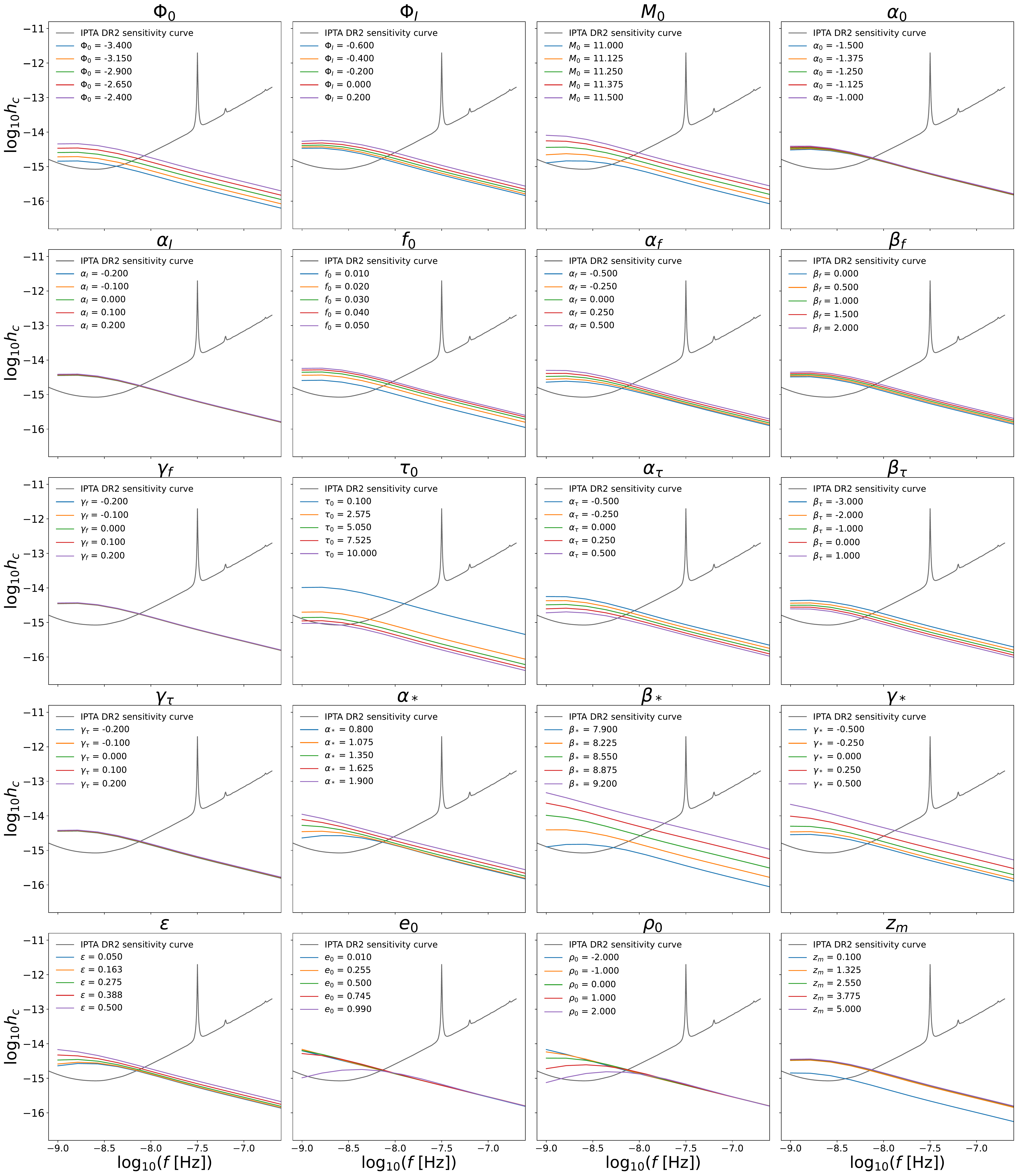}
	\caption{Effect on the GWB spectrum from variations of each of the 20 astrophysical parameters within the range given in \autoref{tab:prior} and \autoref{tab:prior_other}. The default values for the parameters are set as: $\Phi_0 = -2.6, \Phi_I = -0.45, M_0 = 11.25, \alpha_0 = -1.15, \alpha_I = -0.1, f_0 = 0.02, \alpha_f = 0.1, \beta_f = 0.8, \gamma_f = 0.1, \tau_0 = 0.8, \alpha_\tau = -0.1, \beta_\tau = -2., \gamma_\tau = -0.1, \alpha_* = 1.1, \beta_* = 8.2, \gamma_* =-0.2, \varepsilon = 0.3, e_0 = 0.9,  \log_{10}\rho_0 = 0.1, z_m = 2.0$. Each panel shows the change in the GWB spectrum by varying only one parameter.}
	
	\label{fig:gwb_para_variation}
\end{figure*}

\section{Parameter values for the simulated data sets}
\label{sec:appB}
\begin{table*}
	\centering
	\begin{tabular}{| c | c | c| c | c| c | c| c | c| c | c|} 
		\hline
		Parameter & \multicolumn{5}{|c}{Line} &\multicolumn{5}{|c|}{Curve}  \\ [0.5ex] 
		\hline \hline
		$h_c $ at $f=\frac{1}{1\text{year}}$ & $0.5 \times 10^{-15}$ & $1 \times 10^{-15}$ & $2 \times 10^{-15}$ & $3 \times 10^{-15}$ & $4 \times 10^{-15}$ & $0.5 \times 10^{-15}$ & $1 \times 10^{-15}$ & $2 \times 10^{-15}$ & $3 \times 10^{-15}$ & $4 \times 10^{-15}$  \\ [0.5ex] 
		\hline
		$\Phi_0$ & -2.9  & -2.6    & -2.9   & -2.9  & -2.9 & -2.55 & -2.5  & -2.5 & -2.5 & -2.6 \\
 
		$\Phi_I$ & -0.45 & -0.45   & -0.45  & -0.45 & -0.45 & -0.45  & -0.255& -0.1 & 0.08 & 0.095 \\
		$M_0$ & 11.25    & 11.25   &  11.25 & 11.25 &  11.3 &  11.35 & 11.35 & 11.35& 11.35&  11.2 \\ 
		$\alpha_0$ & -1.15 & -1.15 &  -1.15 & -1.15 & -1.15 & -1.1   & -1.1  & -1.1 & -1.1 & -1.1  \\ 
 		$\alpha_I$ & -0.1  & -0.1  & -0.1   & -0.1  & -0.1  & -0.12  & -0.12 & -0.12& -0.12& -0.12 \\ 
		$f_0$      & 0.015 & 0.02  & 0.03   & 0.03  & 0.035 &  0.022 & 0.03  & 0.03 & 0.03 & 0.03  \\ 
		$\alpha_f$ & 0.3   & 0.3   & 0.3    & 0.3   &  0.3  &  -0.15 & -0.15 & -0.15& -0.15& -0.15 \\
		$\beta_f$  & 0.6   & 0.6   & 0.6    & 0.6   &  0.6  &  0.8   & 1.2   & 1.48 & 1.3  &  1.7  \\ 
		$\gamma_f$ & 0.1   & 0.1   & 0.1    & 0.1   &  0.1  &  0.1   & 0.1   & 0.1  &  0.1 & 0.1   \\ 
		$\tau_0$   & 2.    & 1.8   & 2.     & 2.    &  2.   &  0.8   & 0.8   & 0.8  & 0.8  & 0.8   \\ 
		$\alpha_\tau$ & 0.1 & 0.1  & 0.1    & 0.1   &  0.1  &  0.2   & 0.2   & 0.2  &  0.2 & -0.2  \\ 
		$\beta_\tau$ & -1.8 & -2.  & -2.1   & -2.3  & -2.5  &  -2.   & -2.   & -2.  & -2.1 & -2.1  \\ 
		$\gamma_\tau$& -0.1 & -0.1 & -0.1   & -0.1  & -0.1  &  -0.1  & -0.1  & -0.1 & -0.1 & 0.1   \\ 
		$\alpha_*$   & 1.1  & 1.1  & 1.1    & 1.1   &  1.1  &   1.   & 1.    &  1.  &  1.  & 1.    \\ 
		$\beta_*$    & 8.5  & 8.5  & 8.65   & 8.8   &  8.8  &   8.0  & 8.0   & 8.0  &  8.0 & 8.1   \\ 
		$\gamma_*$   & 0.1  & 0.1  & 0.3    & 0.3   &  0.3  &   0.1  & 0.1   & 0.1  &  0.1 & 0.1   \\ 
		$\varepsilon$   & 0.35 & 0.35 & 0.35   & 0.35  &  0.35 &   0.2  & 0.2   & 0.2  &  0.2 & 0.2   \\ 
		$e_0$        & 0.5  &  0.5 & 0.5    & 0.5   &  0.5  &   0.9  & 0.9   & 0.9  &  0.9 & 0.9   \\ 
		$\log_{10}\rho_0$     & -0.1 & -0.1 & -0.1   & -0.1  & -0.1  &  -0.1  & -0.1  & -0.1 & -0.1 & -0.1  \\ 
		$z $         & 1.5  & 2.0  & 3.0    & 3.0   &  3.0  &   1.4  & 2.0   & 3.0  &  3.0 & 3.0   \\ [1ex]
		\hline
	\end{tabular}
	\caption{Values of astrophysical parameters used to create the different simulated detections shown in \autoref{fig:gwb_sim_input}. Note that these are just one possible set for each spectrum and are neither unique nor necessarily representative.}\label{tab:parameter_val}
\end{table*}


\label{lastpage}

\end{document}